\documentclass[showpacs,amsmath,amssymb,aps,showkeys,floatfix,prd,a4paper]{revtex4}

\usepackage[dvips]{graphicx}
\usepackage{dcolumn}
\usepackage{bm}
\usepackage{epsfig}
\usepackage{amsfonts}
\usepackage{amssymb,amscd}

\def\lsim{\raise0.3ex\hbox{$<$\kern-0.75em\raise-1.1ex\hbox{$\sim$}}}
\def\gsim{\raise0.3ex\hbox{$>$\kern-0.75em\raise-1.1ex\hbox{$\sim$}}}

\def\pom{{I\!\!P}}

\def\beq{\begin{equation}}
\def\eeq{\end{equation}}
\def\bea{\begin{eqnarray}}
\def\eea{\end{eqnarray}}
\def\bq{\begin{quote}}
\def\eq{\end{quote}}

\def\gappeq{\mathrel{\rlap {\raise.5ex\hbox{$>$}}
{\lower.5ex\hbox{$\sim$}}}}

\def\lappeq{\mathrel{\rlap{\raise.5ex\hbox{$<$}}
{\lower.5ex\hbox{$\sim$}}}}

\def\Toprel#1\over#2{\mathrel{\mathop{#2}\limits^{#1}}}

\def\pom{{I\!\!P}}
\def\reg{{I\!\!R}}

\begin{document}


\title{Exclusive and diffractive $\mu^+ \mu^-$  production in $pp$ collisions at the LHC}

\author{V. P. Gon\c{c}alves$^{1}$, M. M. Jaime$^{1}$, D. E. Martins$^2$  and  M. S. Rangel$^{2}$ }
\affiliation{$^{1}$ Instituto de F\'{\i}sica e Matem\'atica,  Universidade
Federal de Pelotas (UFPel), \\
Caixa Postal 354, CEP 96010-900, Pelotas, RS, Brazil}

\affiliation{$^{2}$ Instituto de F\'isica, Universidade Federal do Rio de Janeiro (UFRJ), 
Caixa Postal 68528, CEP 21941-972, Rio de Janeiro, RJ, Brazil}

\date{\today}

\begin{abstract}

In this paper we estimate the production of dimuons ($\mu^+ \mu^-$) in exclusive photon -- photon ($\gamma \gamma$) and diffractive Pomeron - Pomeron ($\pom \pom$), Pomeron - Reggeon ($\pom \reg$) and Reggeon - Reggeon ($\reg \reg$) interactions in $pp$ collisions at the LHC energy. The invariant mass, rapidity and tranverse momentum distributions are calculated  using the Forward Physics Monte Carlo (FPMC), which allows to obtain realistic predictions for the dimuon  production with  two leading intact hadrons. In particular, predictions taking into account the CMS and LHCb acceptances are presented. Moreover, the contribution of the single diffraction for the dimuon production also is estimated. Our results demonstrate that the experimental separation of these different mechanisms is feasible. In particular, the events characterized by pairs with large squared transverse momentum are dominated by diffractive interactions, which allows to investigate the underlying assumptions present in the description of these processes. 
\end{abstract}


\maketitle

\section{Introduction}
\label{intro}

The study of exclusive processes, characterized by a low hadronic multiplicity, intact hadrons and rapidity gaps in final state, became a reality in the last years, with experimental results having been obtained in hadronic collisions at the Tevatron, RHIC and LHC \cite{cdf,star,phenix,alice,alice2,lhcb,lhcb2,lhcb3,lhcbconf,exp1,exp2,exp3,exp4,exp5,exp6,exp7}. The study of these processes is mainly motivated by the possibility of improvement of our understanding of the strong interactions theory as well constrain possible scenarios for the beyond Standard Model physics (For a recent review see, e.g. Ref. \cite{review_forward}).  In particular, it is expected that the forthcoming data can be used to discriminate between different approaches for the Quantum Chromodynamics (QCD) at high energies as well as for the  {Pomeron}, which is a long-standing puzzle in the Particle Physics \cite{pomeron}. This object, with the vacuum quantum numbers, is associated with diffractive events, characterized by the presence of large rapidity gaps in the hadronic final state.

\begin{center}
\begin{figure}[t]
\includegraphics[width=0.28\textwidth]{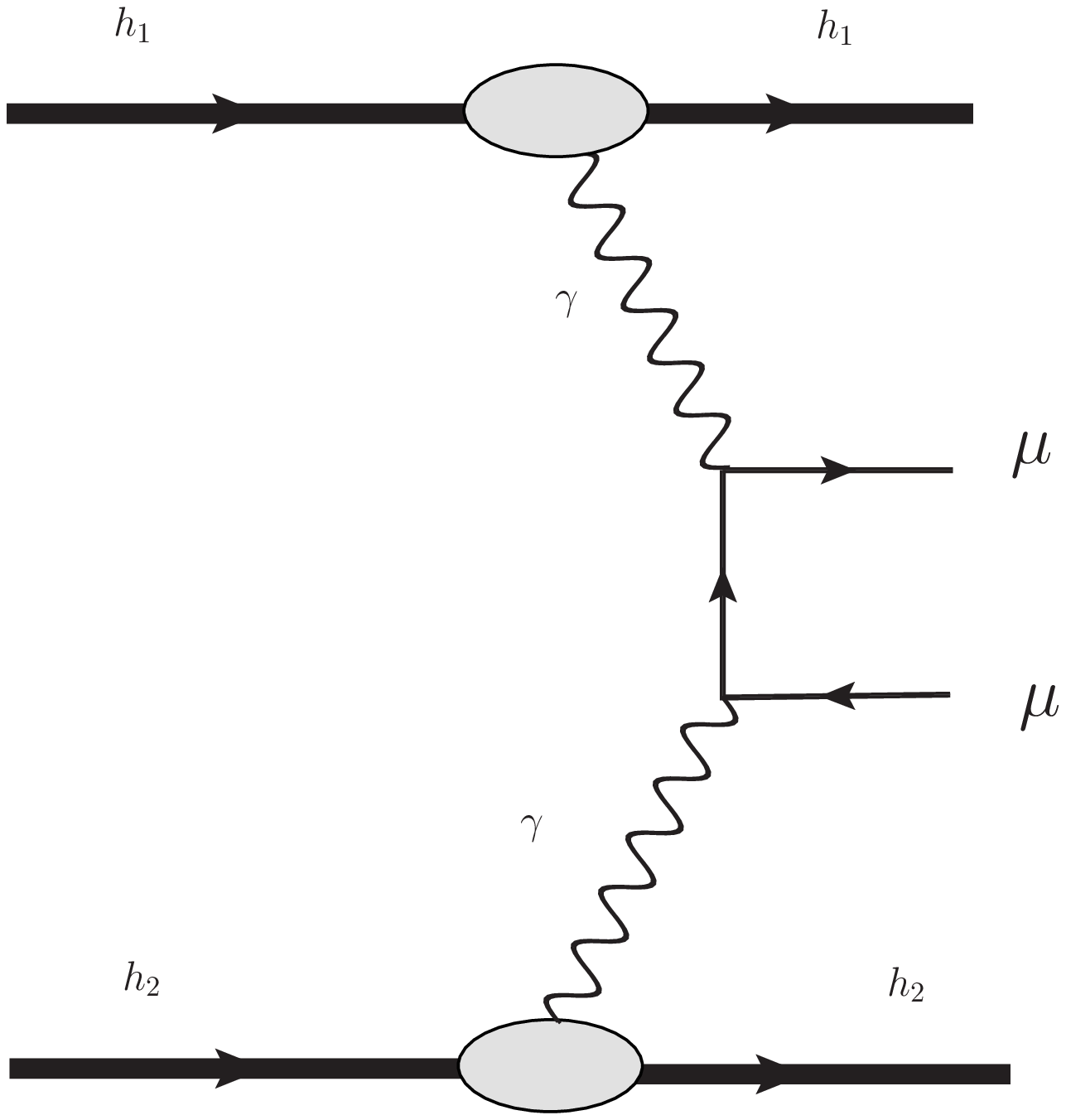}
\includegraphics[width=0.31\textwidth]{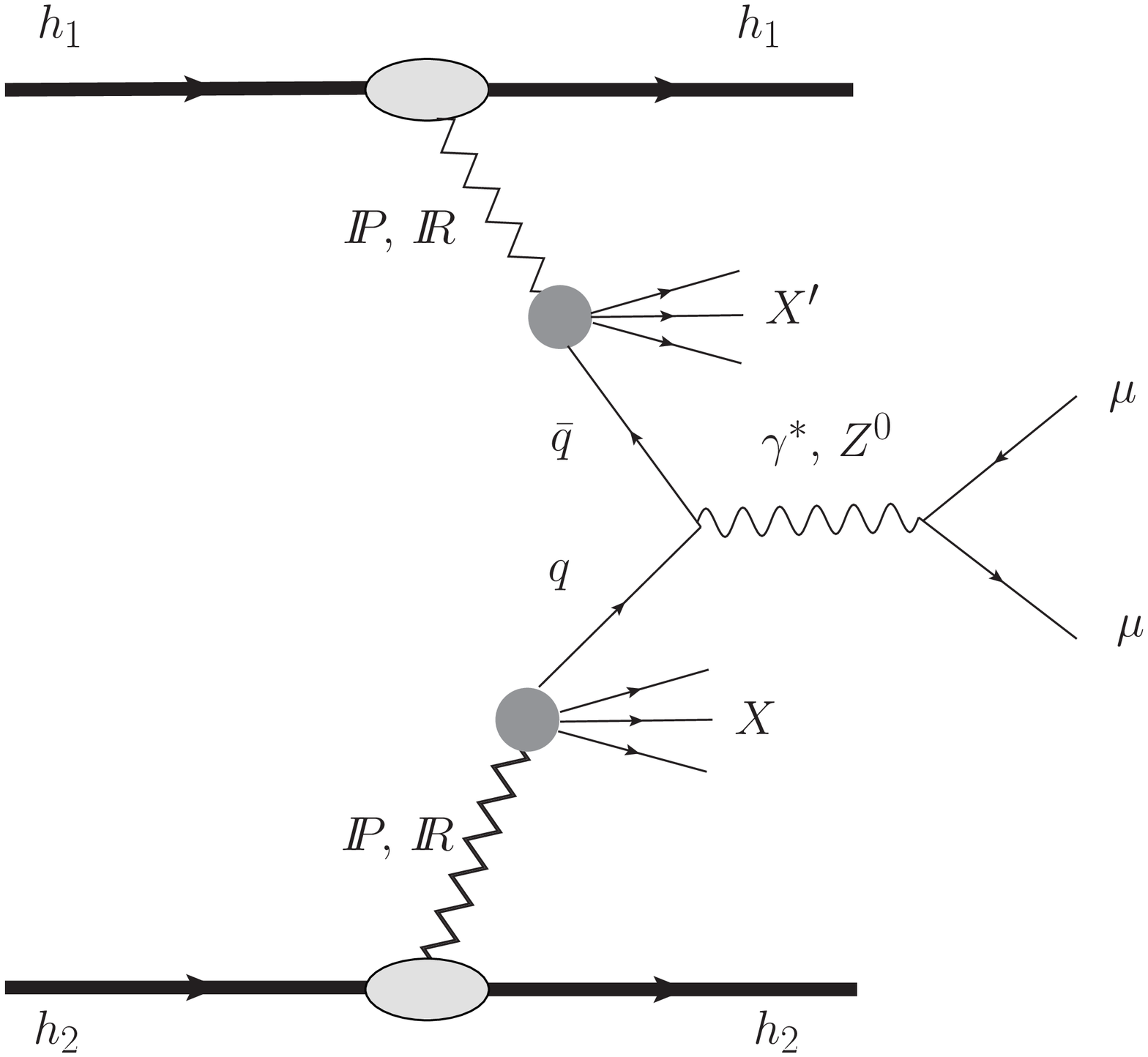}
\includegraphics[width=0.39\textwidth]{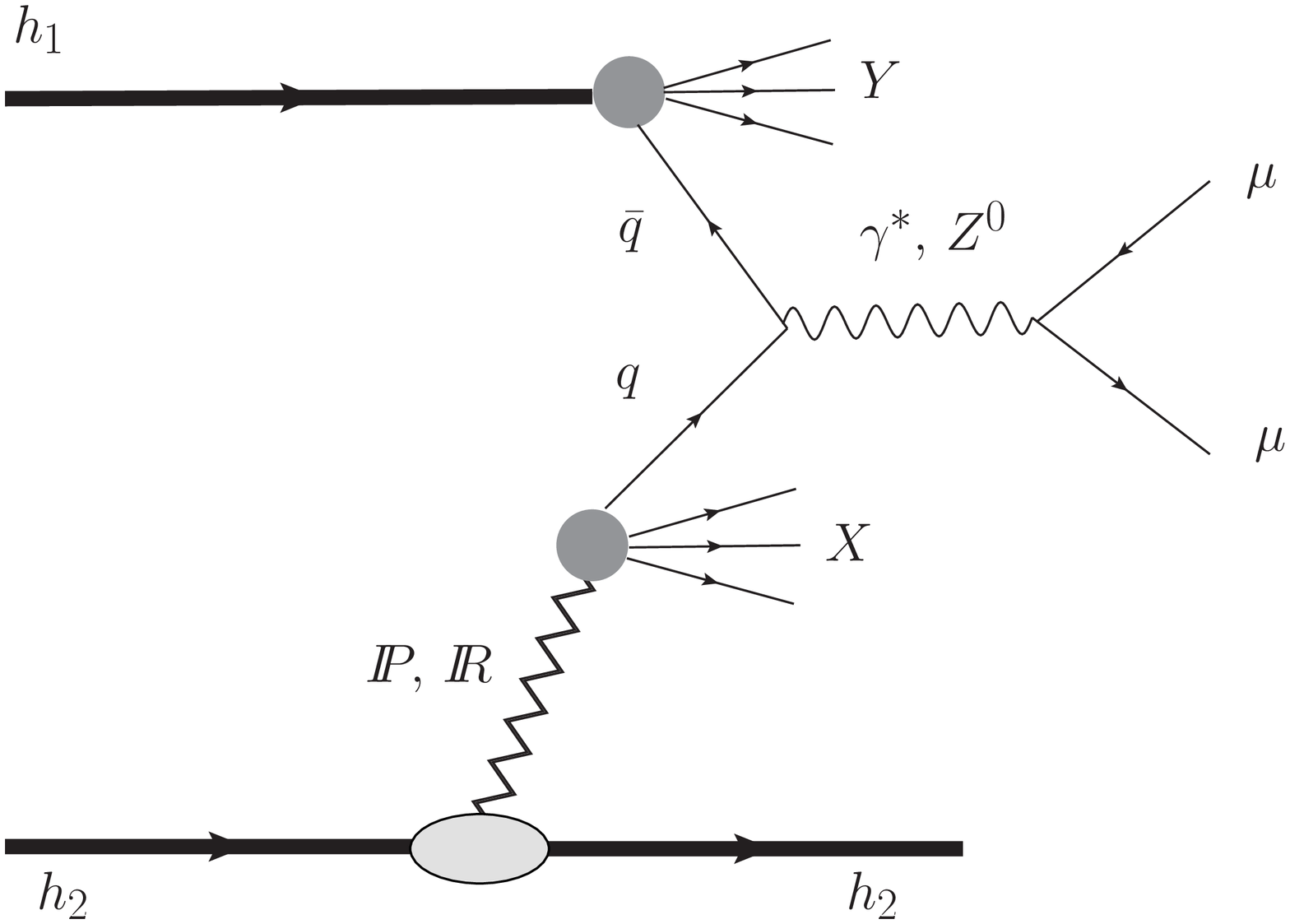}
\caption{The dimuon production in exclusive $\gamma \gamma$ interactions (left panel) and in double (central panel) and single  (right panel) diffractive processes.}
\label{fig:dia}
\end{figure}
\end{center}

One of the more basic examples of an exclusive process is the dimuon ($\mu^+ \mu^-$) production by $\gamma \gamma$ interactions in $pp$ collisions. Such process is considered as being ideal to monitor the collider luminosity \cite{serbo}. During the last years several authors have discussed the backgrounds for this process. In particular, the  contribution of the semielastic and inelastic dimuon production by $\gamma \gamma$ interactions, where one or both incident protons dissociate in the process, have been analyzed in detail in a series of studies (See e.g.\cite{gustavoantoni,vicgus,antoniwolf,khoze}), and important improvements about the treatment of the photon distribution in the proton were derived recently \cite{salam}. Two other backgrounds are the exclusive dimuon production in $\gamma \pom$ interactions \cite{dileptongamapom} and the dimuon production in double diffractive processes \cite{antoniDY}, mediated by Pomeron - Pomeron ($\pom \pom$), Pomeron - Reggeon ($\pom \reg$) and Reggeon - Reggeon ($\reg \reg$) interactions. Both processes,  also   generated two rapidity gaps and two leading protons into the final state. A first comparison between exclusive $\gamma \pom$ and $\pom \pom$ mechanisms was presented in Ref. \cite{antoniDY}, which  demonstrated that both can be larger than the exclusive production in some regions of the phase space. This result strongly motivate the analysis that will be performed in this paper, where we also estimate the $\reg \reg$ and $\pom \reg$ contributions, which becomes significant in some regions of the phase space (For a similar analysis for dijet production see Refs. \cite{danielmurilo,vicmurdijet}).  We will restrict our study to a comprehensive comparison between the exclusive and double diffractive mechanisms for the dimuon production, represented in the left and central panels of Fig. \ref{fig:dia}, respectively. As both processes are implemented in the  Forward Physics Monte Carlo (FPMC) \cite{fpmc}, it allows to obtain realistic predictions for the dimuon  production with  two leading intact hadrons, taking into account the acceptance of the LHC detectors. The inclusion of the exclusive $\gamma \pom$ mechanism in the FPMC is an important task that we intend to do in a near future. In our study, we also will estimate the contribution of the single diffractive dimuon production, represented in Fig. \ref{fig:dia} (right panel),  which is characterized by one rapidity gap and the dissociation of one of the incident protons. As currently the LHC experiments are only able to detect one of the outgoing hadrons, in order to select the exclusive events of interest is fundamental to have control of the single diffractive events. Our goal is to identify the regions of dominance of the different mechanisms as well to determine the typical cutoffs that should be applied in order to  establish the clear dominance of the  exclusive and diffractive processes.

The content of this paper is organized as follows. In the next section we present a brief review of the formalism for the dimuon production in photon - photon and \,diffractive  interactions in hadronic collisions. In Section \ref{results} we present our predictions for the invariant mass,  rapidity and  transverse momentum distributions  for the dimuon production in  \,{$pp$} collisions at LHC energy, considering the contributions associated to $\gamma \gamma$, $\pom \pom$,  $\pom \reg$, {$\reg \reg$,  $\pom p$ and $\reg p$} interactions. Finally, in Section \ref{conc} we summarize our main conclusions.

\section{Formalism}
In this Section we will present a brief review of the main concepts needed to describe the $\mu^+ \mu^-$ production by photon - photon and diffractive interactions in $pp$ collisions at the LHC energies.  Initially, let's consider the exclusive $\mu^+ \mu^-$  production by $\gamma \gamma$ interactions in the collision of two hadrons, $h_{1}$ and $h_{2}$, represented in Fig. \ref{fig:dia} (left panel). In the equivalent photon approximation \cite{upc},  the cross section    is given by 
\begin{eqnarray}
\sigma(h_{\rm 1} h_{\rm 2} \rightarrow h_{\rm 1} \otimes \mu^+ \mu^- \otimes h_{\rm 2}) = \int dx_{\rm 1} \int dx_{\rm 2} \, \gamma_{\rm 1}(x_{\rm 1},Q^2) \cdot \gamma_{\rm 2}(x_{\rm 2},Q^2) \cdot \hat{\sigma}(\gamma \gamma \rightarrow \mu^+ \mu^-) \,\,,
\label{fotfot}
\end{eqnarray}
where $\gamma_i(x_i,Q^2)$ is the equivalent photon distribution of the hadron $i$, with  $x_i$ being the fraction of the hadron energy carried by the photon and $Q^2$ has to be identified with a hard scale of the  process. Moreover, $\otimes$ represents the presence of a rapidity gap in the final state and 
$\hat{\sigma}$ is the  cross section for the  $\gamma \gamma \rightarrow \mu^+ \mu^-$ process. The basic idea of the Eq. (\ref{fotfot}) is that at high energies,  {an} ultra relativistic proton
 give rise to strong electromagnetic fields, such that the photon stemming from the electromagnetic field of one of the two colliding protons can interact with a photon of the other proton, and generate a given final state \cite{upc,epa}. The  ingredients for the analysis of the exclusive dimuon production by $\gamma \gamma$ interactions are the elementary cross section $\hat{\sigma}$, which is well known from QED, and  the  equivalent photon distribution of the incident protons. In the case of of a charged  \,{\it point-like} fermion, the equivalent photon distribution was formulated  many years ago by Fermi \cite{Fermi} and developed by Williams \cite{Williams} and Weizsacker \cite{Weizsacker}.  On the other hand,  the calculation of the photon distribution of the proton have been estimated by several authors \cite{epa,kniehl,dz} considering different approximations. One of the more detailed derivations have been presented by Ginzburg and collaborators in Ref. \cite{epa}, where an analytical expression have been derived, which will be used in our further calculations. As demonstrated in Ref. \cite{vicwerdaniel} the difference between the different modellings of the photon flux is smaller than 5\% at low-$x$. 
 
Let's discuss now the dimuon production in diffractive processes, represented in the central and right panels of Fig. \ref{fig:dia}. Assuming the validity of the factorization theorem, the cross section for the {\it double} diffractive $\mu^+ \mu^-$ production can be expressed by  
\begin{eqnarray}
\sigma(h_{\rm 1} h_{\rm 2} \rightarrow h_{\rm 1} \otimes X \mu^+ \mu^- X^{\prime} \otimes h_{\rm 2}) = \int dx_{\rm 1} \int dx_{\rm 2} &[& q^D_{\rm 1}(x_{\rm 1},Q^2) \cdot \bar{q}^D_{\rm 2}(x_{\rm 2},Q^2) +  \nonumber \\ &&\bar{q}^D_{\rm 1}(x_{\rm 1},Q^2) \cdot {q}^D_{\rm 2}(x_{\rm 2},Q^2)\, ] \cdot \hat{\sigma}(q \bar{q} \rightarrow \mu^+ \mu^- ) \,\,,
\label{pompom}
\end{eqnarray}
where $q^D_i (x_i,Q^2)$ and $\bar{q}^D_i (x_i,Q^2)$ are the   diffractive quark and antiquark distributions of the hadron $i$ with a momentum fraction $x_i$ and $\hat{\sigma}(q \bar{q} \rightarrow \mu^+ \mu^- )$ is the cross section for the Drell - Yan process. Similarly, the cross section for the {\it single} diffractive $\mu^+ \mu^-$ production is given by   
\begin{eqnarray}
\sigma(h_{\rm 1} h_{\rm 2} \rightarrow  Y \mu^+ \mu^- X \otimes h_{\rm i} ) = \int dx_{\rm 1} \int dx_{\rm 2} \, [ q^D_{\rm 1}(x_{\rm 1},Q^2) \cdot \bar{q}_{\rm 2}(x_{\rm 2},Q^2) &+&\nonumber\\  q_{\rm 1}(x_{\rm 1},Q^2) \cdot \bar{q}^D_{\rm 2}(x_{\rm 2},Q^2) &+&  (q \leftrightarrow \bar{q})] \cdot \hat{\sigma}( q \bar{q} \rightarrow \mu^+ \mu^-) \,\,, 
\label{pompho}
\end{eqnarray}
where $h_i$ represents the hadron that have emitted the Pomeron and remains intact. Moreover,  $q_{\rm i}(x_{\rm i},Q^2)$ and $\bar{q}_{\rm i}(x_{\rm i},Q^2)$ are the standard inclusive quark and antiquark distributions of the proton. The inclusive and diffractive parton distributions are non-perturbative objects. However, the evolution with the factorization scale $Q^2$ is described perturbatively by the DGLAP evolution equations.  
 In our calculations we will assume that the factorization scale is  the square of the dimuon invariant mass, i.e., $Q^2 = M^2_{\mu^+ \mu^-}$.
  In order to describe the diffractive parton distributions we will consider in what follows the Resolved Pomeron model \cite{IS}, which implies that these quantities can be expressed in terms of the Pomeron ($\pom$) and Reggeon ($\reg$) contributions as follows
\begin{eqnarray}
{ q^D_{p}(x,Q^2)}= { \int_x^1 \frac{d\xi}{\xi} f^{p}_{\pom}(\xi) ~q_{\pom}\left(\frac{x}{\xi}, Q^2\right)} + { \int_x^1 \frac{d\xi}{\xi} f^{p}_{\reg}(\xi) ~q_{\reg}\left(\frac{x}{\xi}, Q^2\right)}  \,\,,
\label{difquark:proton}
\end{eqnarray}
where $\xi$ is the momentum fraction of the proton carried by the Pomeron and Reggeon, $f^{p}_{\pom,\,\reg}(\xi)$ are the associated flux distributions in the proton and  
$q_{\pom,\,\reg}(\beta \equiv {x}/\xi , Q^2)$ are its corresponding  quark distributions. Moreover,  $\beta $ is the momentum fraction carried by the partons inside the \,{Pomeron} and Reggeon. 
Following Ref. \cite{H1diff}, we assume that the  \,{Pomeron} and Reggeon fluxes are given by
\begin{eqnarray}
f^{p}_{\pom}(\xi)= 
\int_{t_{\rm min}}^{t_{\rm max}} dt \, \frac{A_{\pom} \, e^{B_{\pom} t}}{\xi^{2\alpha_{\pom} (t)-1}}  \,\,\,\,\,\mbox{and} \,\,\,\,\,f^{p}_{\reg}(\xi)= 
n_{\reg}\cdot \int_{t_{\rm min}}^{t_{\rm max}} dt \, \frac{A_{\reg} \, e^{B_{\reg} t}}{\xi^{2\alpha_{\reg} (t)-1}}  \,\,,
\label{fluxpom:proton}
\end{eqnarray}
where $t_{\rm min}$, $t_{\rm max}$ are kinematic boundaries and $n_{\reg}$ is a normalization factor for the Reggeon term. The flux factors are motivated by Regge theory, where the \,{pomeron} and reggeon trajectories are assumed to be linear, $\alpha_{\pom,\reg} (t)= \alpha_{\pom,\reg} (0) + \alpha_{\pom,\reg}^\prime t$, and the parameters $B_{\pom,\reg}$, $\alpha_{\pom,\reg}^\prime$, $n_{\reg}$ and their uncertainties are obtained from fits to H1 data  \cite{H1diff}. As demonstrated in Ref. \cite{H1diff}, the HERA data is able to contrain the Pomeron structure, which is dominated by gluons. However, a Reggeon contribution is required to describe the experimental data at large $\xi$. As in Ref. \cite{H1diff}, we will assume that the Reggeon contribution can be modelled by a quark - antiquark exchange and its structure can be described in terms of the pion structure function. In general, the Reggeon contribution is disregarded in the diffractive calculations of different final states. However, as demonstrated in Ref. \cite{danielmurilo}, such contribution can be important in some regions of the phase space. In the next Section we will estimate, by the first time, the impact of $\reg \reg$ and $\pom \reg$ interactions for the dimuon production.

One important open question in the treatment of exclusive and diffractive interactions in hadronic collisions is if the cross sections for the associated processes are not somewhat modified by soft interactions which lead to an extra production of particles that destroy the rapidity gaps in the final state \cite{bjorken}. As these effects have nonperturbative nature, they are difficult to treat and its magnitude is strongly model dependent (For recent reviews see Refs. \cite{durham,telaviv}). In the case of diffractive interactions in \,{${pp / p\bar{p}}$} collisions, the experimental results obtained at TEVATRON \cite{tevatron} and LHC \cite{atlas_dijet,cms_dijet} have demonstrated that   one should take into account of these additional absorption effects that imply the violation of the QCD hard scattering factorization theorem for diffraction \cite{collinsfac}. In general, these   effects are parametrized in terms of a rapidity gap survival probability, $S^2$, which corresponds to the probability of the scattered proton not to dissociate due to the secondary interactions. Different approaches have been proposed to calculate these effects  giving distinct predictions (See, e.g. Ref. \cite{review_martin}). An usual approach in the literature is the calculation of an average probability $\langle |S|^2\rangle$ and after to multiply  the cross section by this value. As previous studies for the double diffractive production 
\cite{nosbottom,MMM1,antoni,antoni2,cristiano,cristiano2,kohara_marquet} we also follow this simplified approach assuming $\langle |S|^2\rangle = 0.02$ for the double diffractive  production and $\langle |S|^2\rangle = 0.05$ for the single diffractive one. It is important to emphasize that this choice is somewhat arbitrary, and mainly motivated by the possibility to compare our predictions with those obtained in other analysis. Recent studies from the CMS Collaboration \cite{cms_dijet} indicate that this factor can be larger  than this value by a factor $\approx 4$. Consequently, our results can be considered a lower bound for the diffractive contribution.
In the case of $\gamma \gamma$,  we will assume  $\langle |S|^2\rangle = 1$. However, it is important to emphasize that the magnitude of the rapidity gap survival probability in $\gamma \gamma$ still is an open question and some authors proposed that it smaller than the unity \cite{khoze,schoffel}. Therefore,  the results for the dimuon production by $\gamma \gamma$ interactions  may be considered an upper bound.

\section{Results}
\label{results}

In what follows we present our results for the dimuon production  by photon -- photon, Pomeron -- Pomeron and Pomeron -- Reggeon   interactions in \,{$pp$} collisions at the Run 2  LHC  energy (For a similar analysis for the heavy quark and dijet production see Refs. \cite{antoni,nosbottom,vicmurdijet}). 
As discussed in the Introduction, these processes are characterized by two rapidity gaps and intact hadrons in the final state. The experimental separation of these events using the two rapidity gaps to tag the event  is not an easy task  at the LHC  due to the non - negligible \,{pile-up} present in the normal runs. An alternative  is the detection of the outgoing intact hadrons. Recently, the 
ATLAS, CMS and TOTEM Collaborations have proposed the setup of forward detectors \cite{Albrow:2008pn,ctpps,marek}, which will enhance the kinematic coverage for such investigations. Moreover, the LHCb experiment can study diffractive events by requiring forward regions void of particle production $5.5<|\eta|<8.0$~\cite{Jpsi13tev}.

\begin{center}
\begin{figure}[t]
\includegraphics[width=0.32\textwidth]{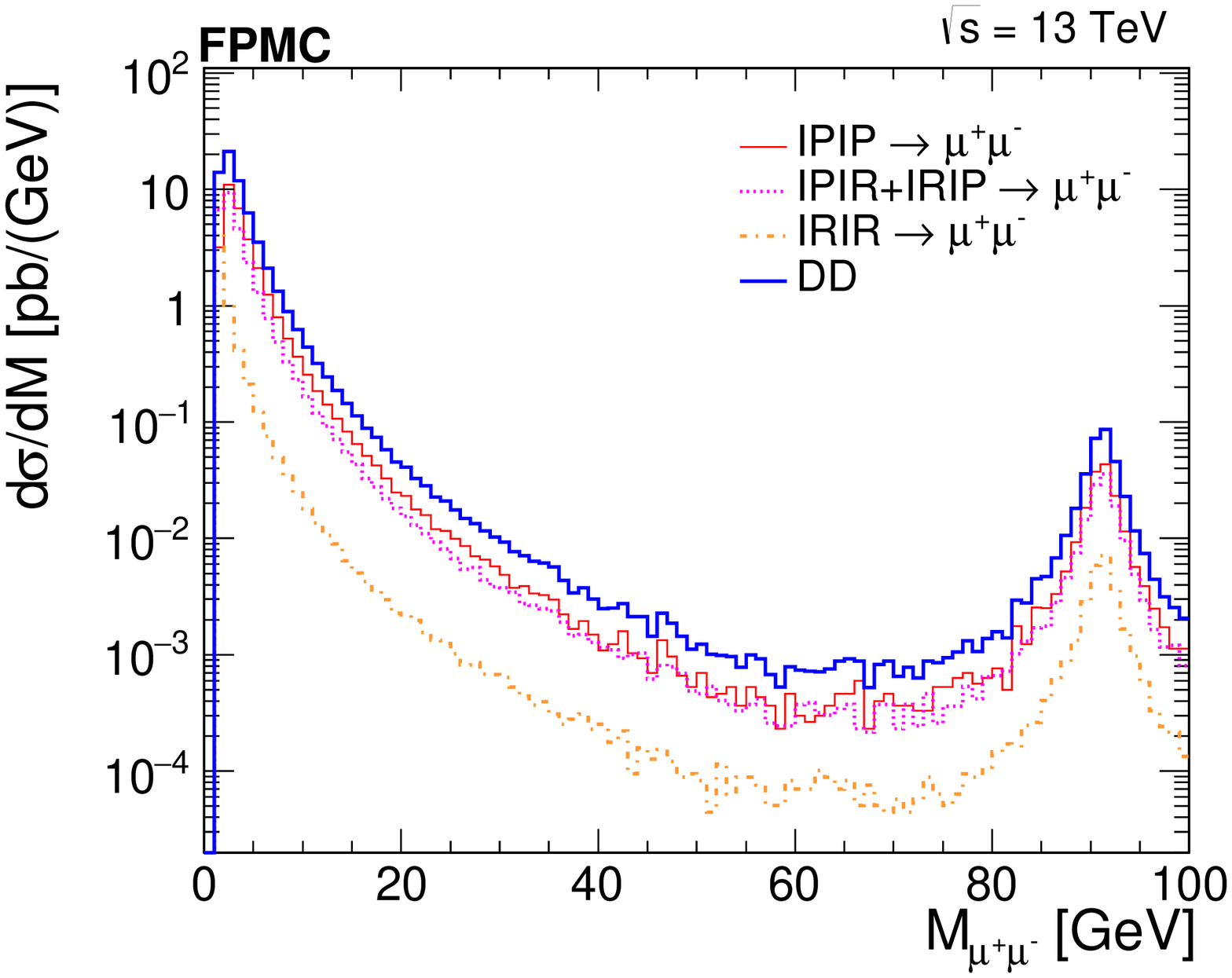}
\includegraphics[width=0.32\textwidth]{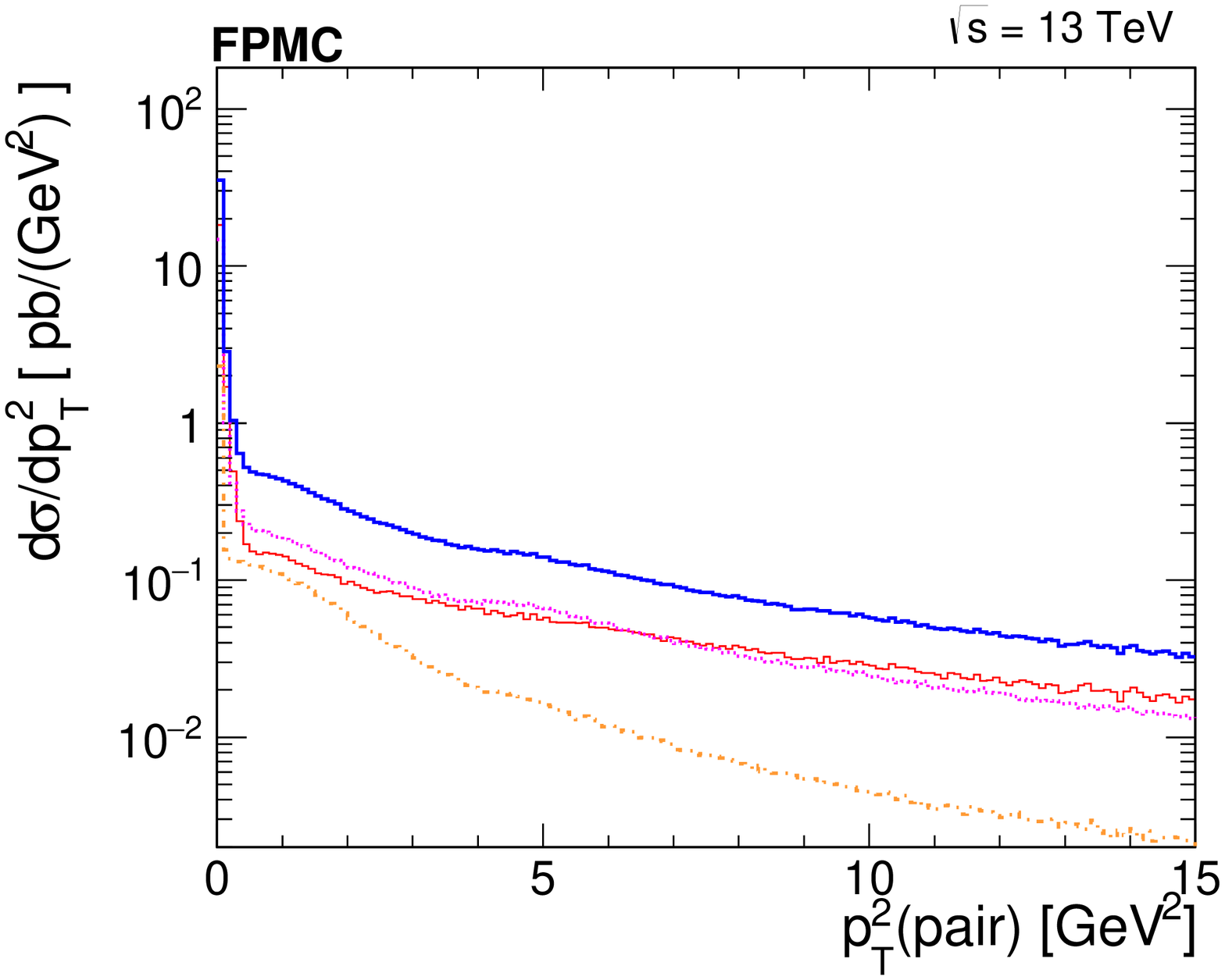}
\includegraphics[width=0.32\textwidth]{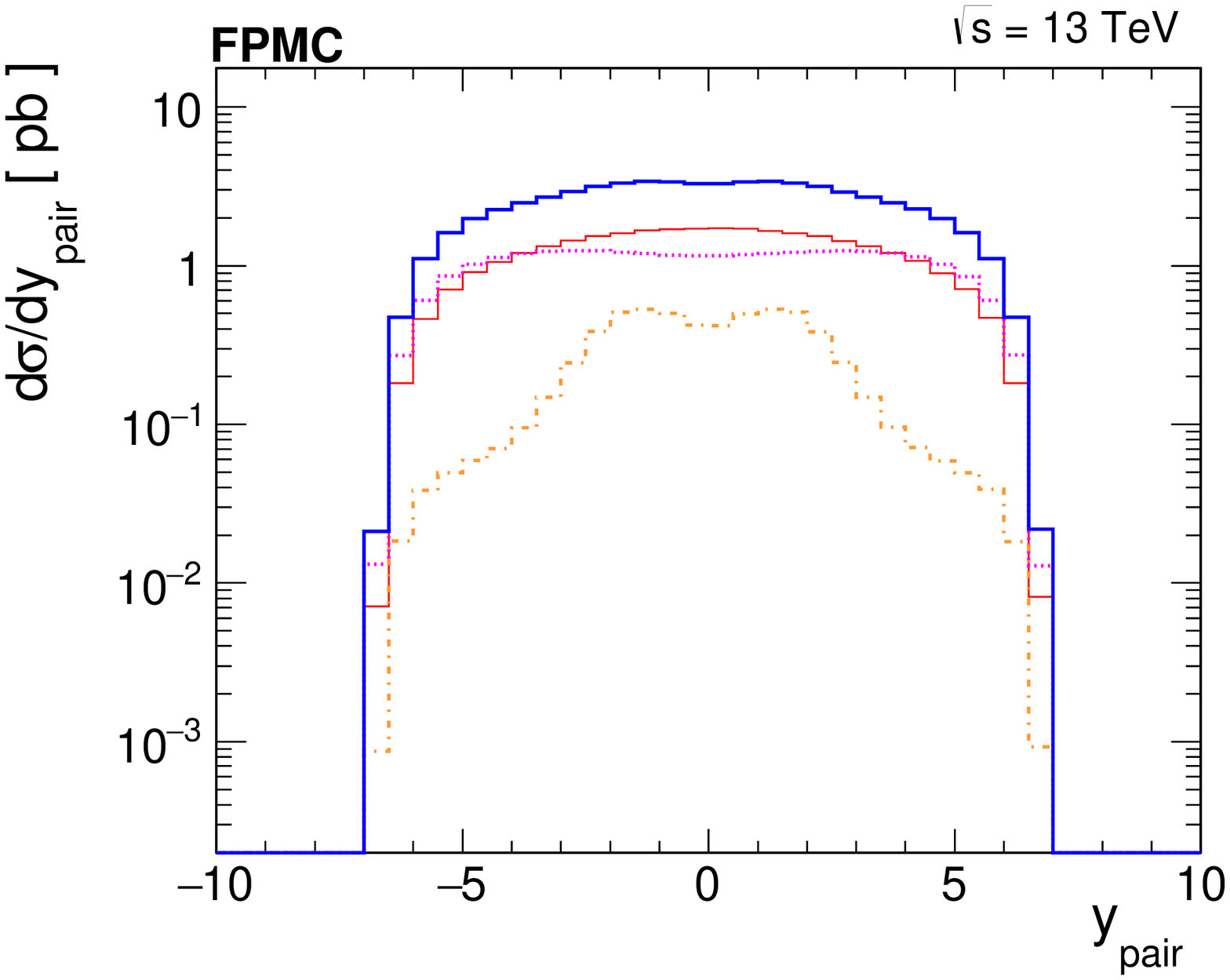}
\caption{ Contribution of the $\pom \pom$, $\pom \reg$ and $\reg \reg$ mechanisms for the invariant mass (left panel), squared pair transverse momentum  (central panel) and pair rapidity (right panel) distributions considering the double diffractive $\mu^+ \mu^-$ production. The sum of the contributions, denoted by DD,  is represented by solid blue line.}
\label{fig:componentsDD}
\end{figure}
\end{center}

In our analysis we will assume {$pp$} collisions at $\sqrt{s} = 13$ TeV. Moreover,  in order to ensure the validity of perturbative calculations we will impose a cut on the dimuon invariant mass $M_{\mu^+ \mu^-} > 1$ GeV.
The cross sections for the $\gamma \gamma \rightarrow \mu^+ \mu^-$ and $q \bar{q} \rightarrow \mu^+ \mu^-$ subprocesses are calculated at leading order in FPMC using HERWIG 6.5. We will assume the CT10 parametrization \cite{ct10} for the standard inclusive parton distributions and the Fit B, provided  by the H1 Collaboration in Ref. \cite{H1diff}, for the diffractive parton distributions. Finally, the subleading contribution for the diffractive interactions associated to the Reggeon exchange will be taken into account.

\begin{center}
\begin{figure}[t]
\hspace*{-2.05em}\includegraphics[width=0.33\textwidth]{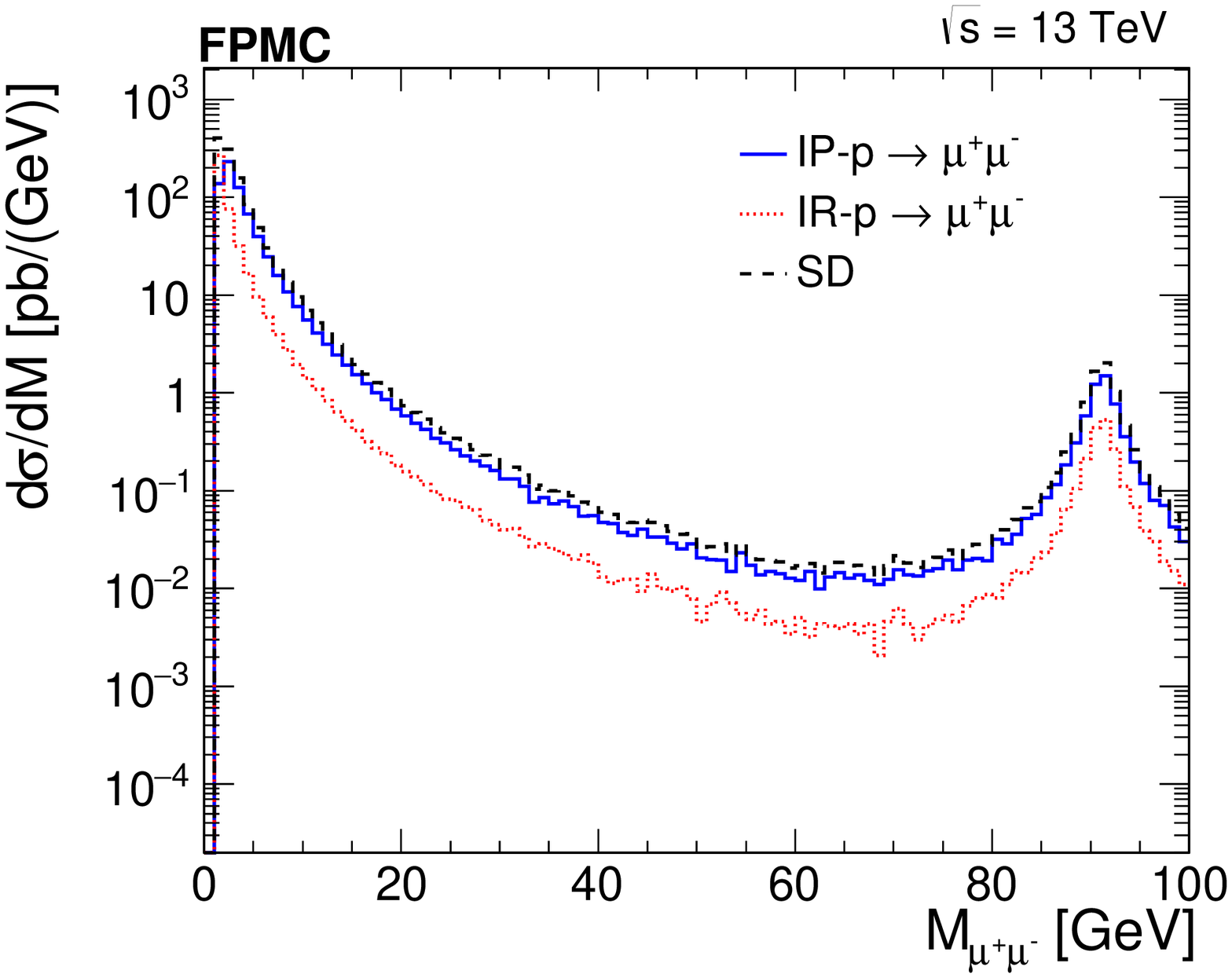}
\hspace*{-0.10em}\includegraphics[width=0.33\textwidth]{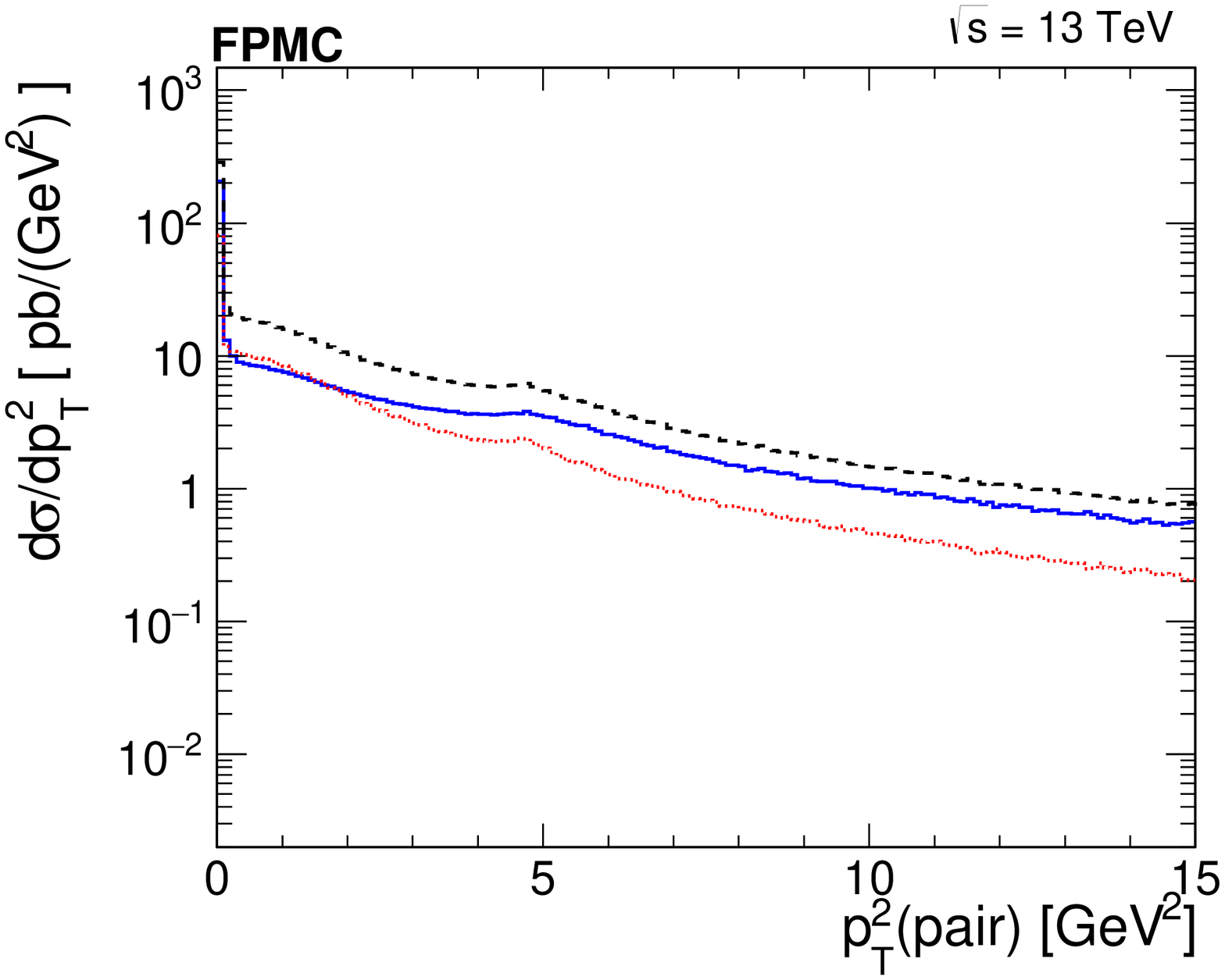}
\hspace*{-0.10em}\includegraphics[width=0.33\textwidth]{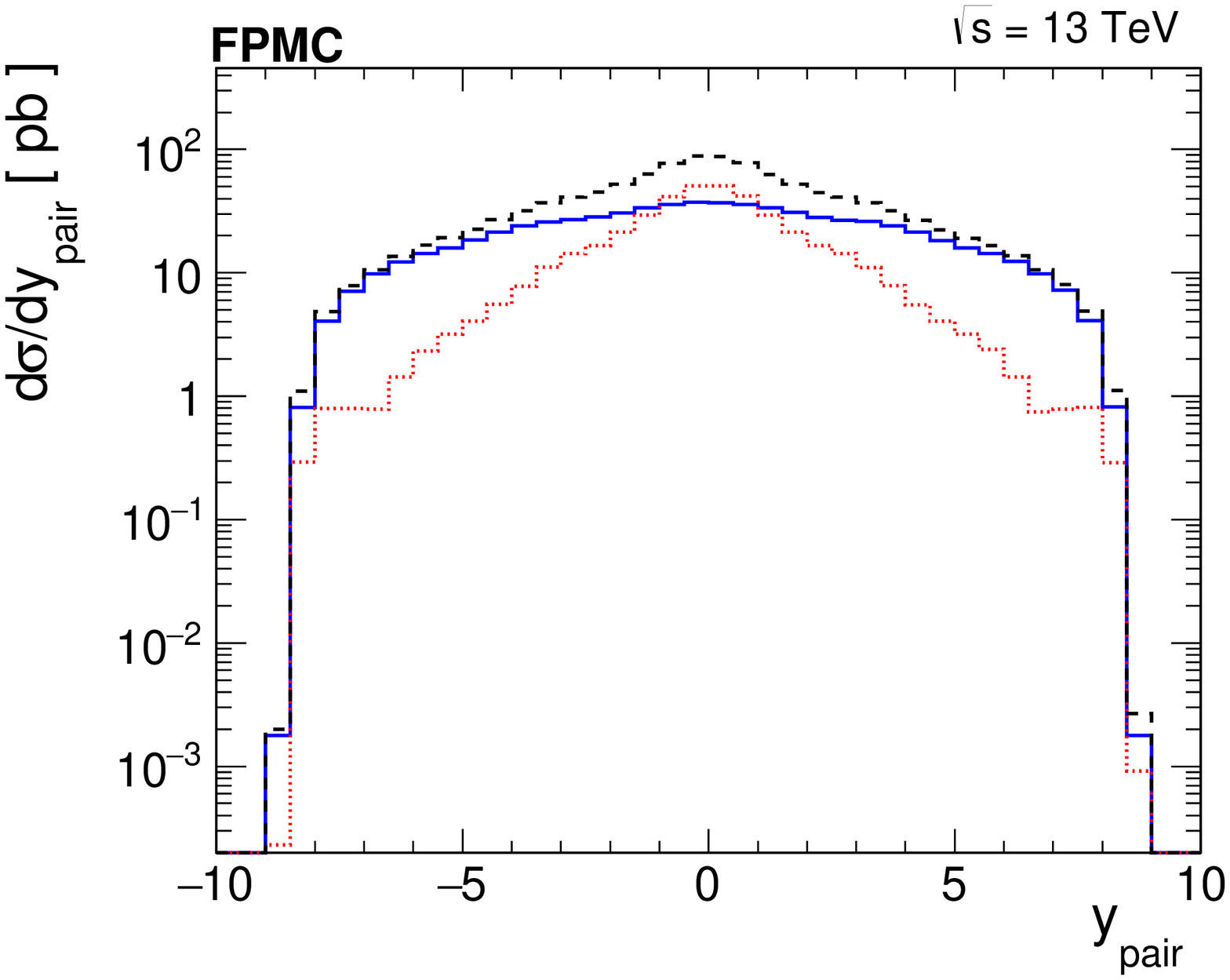}
\caption{ Contribution of the $\pom p$ and $\reg p$ mechanisms for the invariant mass (left panel), squared pair transverse momentum  (central panel) and pair rapidity (right panel) distributions considering the single diffractive $\mu^+ \mu^-$ production. The sum of the contributions, denoted by SD,  is represented by the {dashed black} line. }
\label{fig:componentsSD}
\end{figure}
\end{center}

Initially, let's estimate the contribution of the $\pom \pom$, $\pom \reg$ and $\reg \reg$ mechanisms for the double diffractive $\mu^+ \mu^-$ production. Our results for the invariant mass ($M_{\mu^+ \mu^-}$), squared transverse momentum of the pair ($p_T^2$) and the pair rapidity ($y$) are presented in Fig. \ref{fig:componentsDD}. We have that the $\reg \reg$ contribution is subleading  in the kinematical range considered. However, the $\pom \reg$ one is non-negligible and cannot be disregarded, specially at large values of $y$ and small values of $p_T^2$. In the case of the single diffractive $\mu^+ \mu^-$ production, the basic mechanisms that contribute are associated to $\pom p$ and $\reg p$ interactions, where $p$ indicates that the quark/antiquark comes from the proton instead of the Pomeron or Reggeon. The contribution of these mechanisms are presented in Fig. \ref{fig:componentsSD}. In this case we have that the Reggeon - proton interactions become important at small $p_T^2$ and central rapidities ($y \approx 0$). The results presented in Figs. \ref{fig:componentsDD} and \ref{fig:componentsSD} indicate that the Reggeon contributions are non - negligible for the dimuon production and should be taken into account in order to obtain a more realistic prediction. Such result is expected, since the DY production is associated to the $q \bar{q} \rightarrow \mu^+ \mu^-$ subprocess and the Reggeon structure is dominated by quarks and antiquarks.

\begin{center}
\begin{figure}[t]
\hspace*{-2.05em}\includegraphics[width=0.33\textwidth]{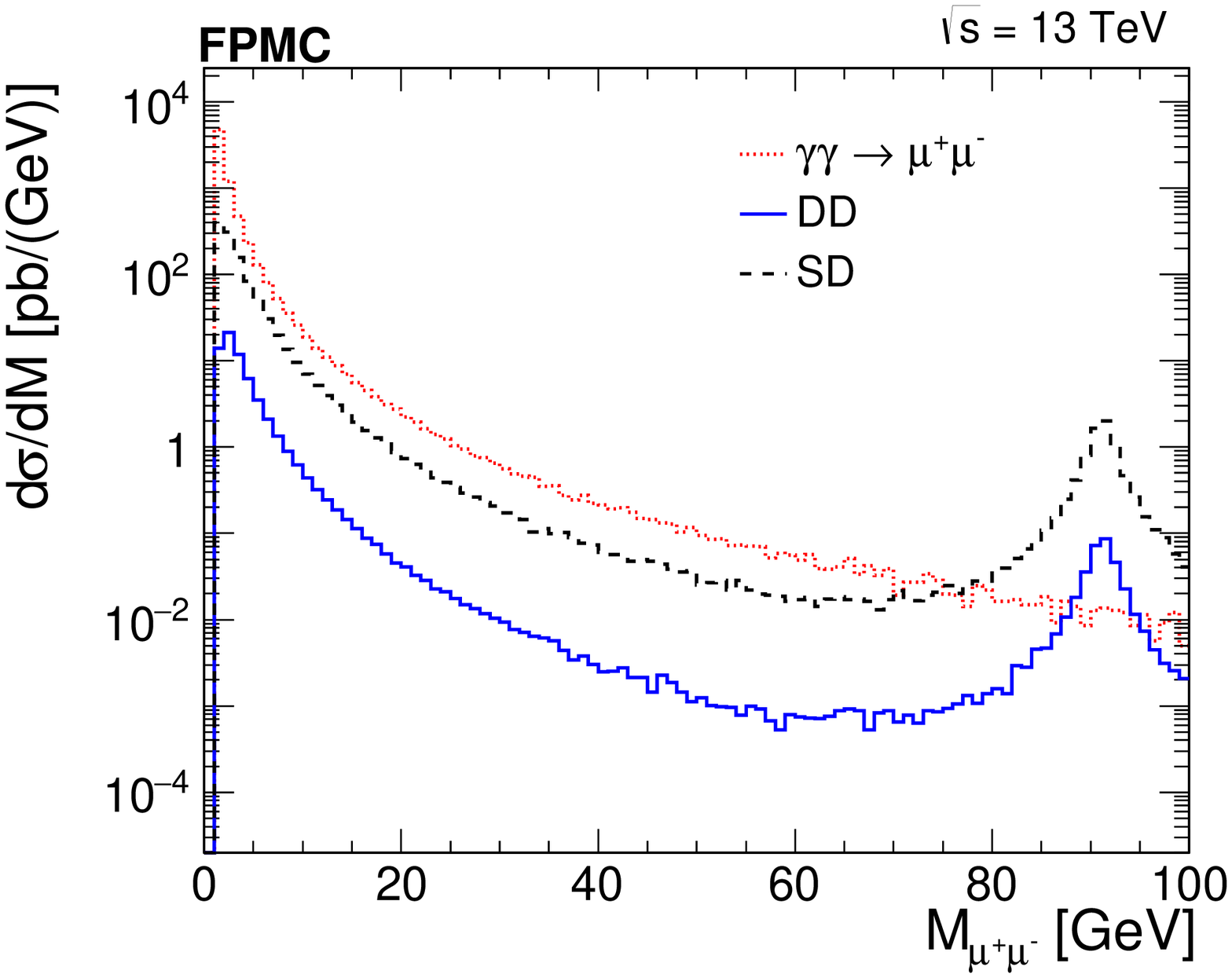}
\hspace*{-0.10em}\includegraphics[width=0.33\textwidth]{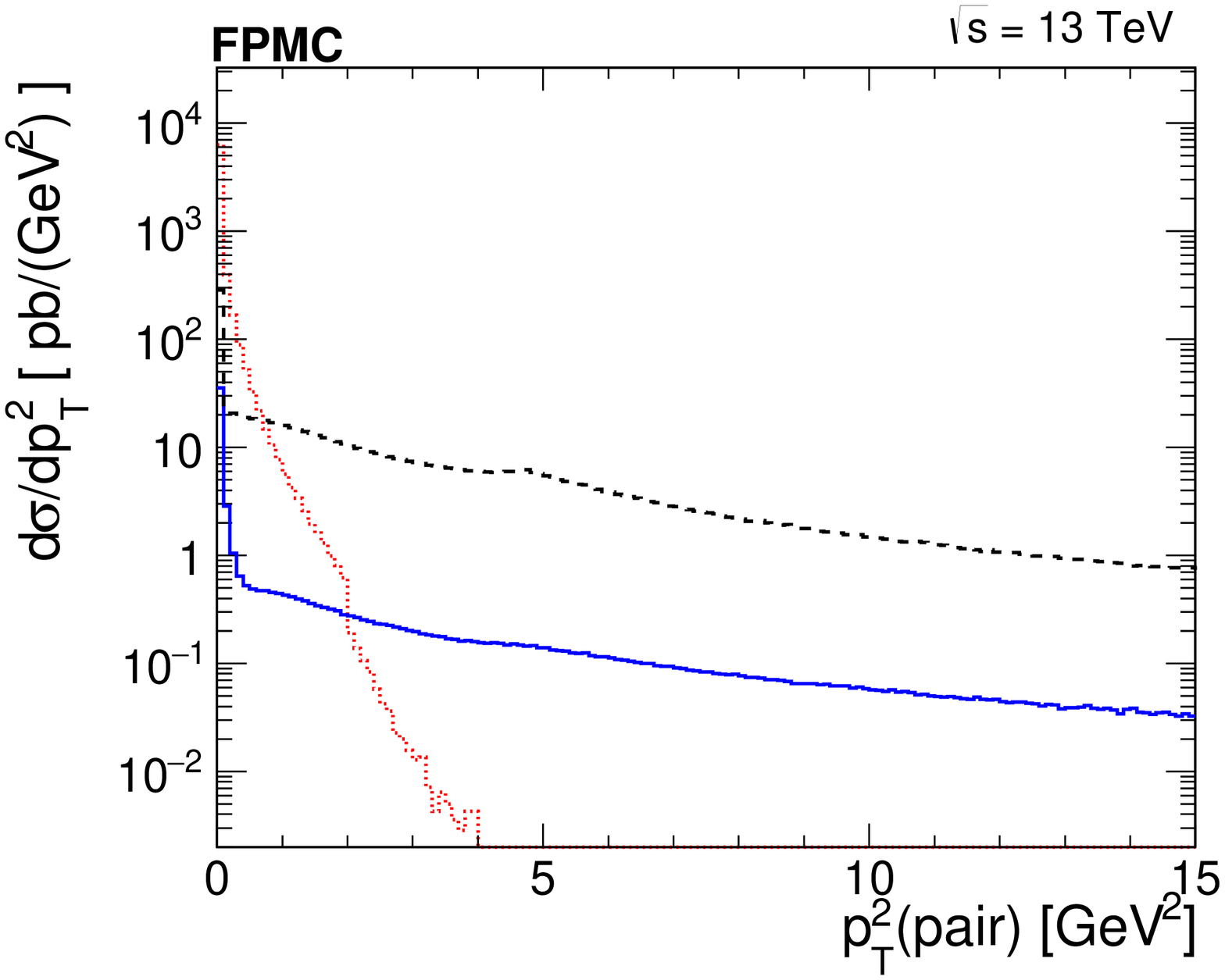}
\hspace*{-0.10em}\includegraphics[width=0.33\textwidth]{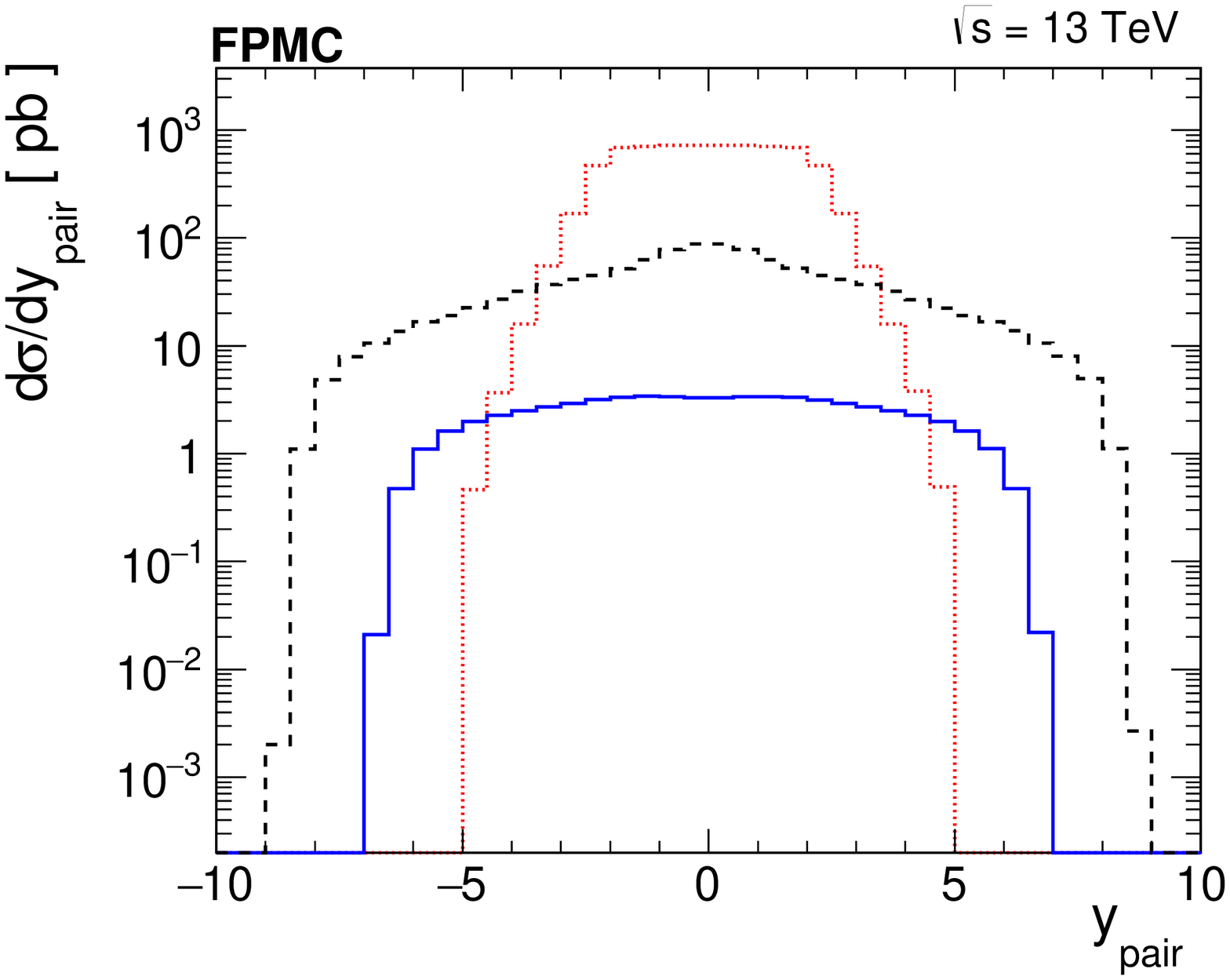}
\caption{Comparison between the predictions for the exclusive, single and double diffractive dimuon production.}
\label{fig:nocuts}
\end{figure}
\end{center}

\begin{center}
\begin{table}[t]
\begin{tabular}{l||c|c|c||c||c|c||c||c }
\hline
\hline
Process & $\mathbb{P}\mathbb{P}$ & $\mathbb{PR}$ + $\mathbb{RP}$ & $\mathbb{RR}$ & {\bf DD} & $\mathbb{P}\boldsymbol{p}$ &
$\mathbb{R}\boldsymbol{p}$ & {\bf SD} & {\bf $\gamma\gamma$}\tabularnewline
\hline
\begin{tabular}{l}
{\footnotesize{}Total Cross Section {[} pb {]}}\tabularnewline
\end{tabular} & {\footnotesize{} 31.0 } & {\footnotesize{} 27.0 } & {\footnotesize{} 6.1 } & {\footnotesize{} 64.1} &
{\footnotesize{} 694.0 } & {\footnotesize{} 425.0 } & {\footnotesize{} 1119.0 } & {\footnotesize{} 7101.1 }\tabularnewline
\hline
\hline
\end{tabular}
\caption{Predictions for the total cross sections  of the different mechanisms for the exclusive and diffractive dimuon production obtained assuming that $M_{\mu^+ \mu^-} > 1$ GeV.}
\label{table:XSeC}
\end{table}
\end{center}

Let's now compare our predictions for the single and double diffractive dimuon production with those for the exclusive $\mu^+ \mu^-$ production by $\gamma \gamma$ interactions. The results are presented in Fig. \ref{fig:nocuts}, where SD and DD indicate the sum of the different mechanisms for the single and double diffractive dimuon production, respectively. We have that the SD and DD mechanisms become dominant for invariant masses close to the $Z^0$ peak, large values of $p_T^2$ and/or $y$.         
 The dominance of the diffractive processes at large $p_T^2$ is expected. As discussed in Refs. \cite{antoniDY,vicgustavo}, the muons produced by $\gamma \gamma$ interactions are emitted preferentially back - to - back, 
with their transverse momenta almost cancelling each other, which implies that the transverse momentum of the pair should be small. On the other hand, in the diffractive mechanisms, the pomeron exchange implies that the typical transverse momentum will be larger \cite{vicgustavo,vicdiego}. Therefore, the exclusive contribution is expected to dominate at small transverse momenta of the pair, while the diffractive one should dominate at large transverse momenta. The corresponding predictions for the total cross sections are presented in Tab. \ref{table:XSeC}. In agreement with our previous discussion, we have that the $\pom \pom$ and $\pom \reg$ contributions for the double diffractive $\mu^+ \mu^-$ production are similar. {Similarly}, the $\reg p$ one is no - negligible for the single diffractive production. In comparison to the exclusive $\mu^+ \mu^-$ production by $\gamma \gamma$ interactions, we {have} that the SD contribution is a factor $\approx 7$ smaller. On the other hand, the DD one is smaller than the exclusive mechanism by two orders of magnitude. As already emphasized in the Introduction, the topology of these distinct mechanisms is different. In particular, in single diffraction, only one rapidity gap is present in the final state. Moreover, in the case of diffractive processes, the presence of remnants of the Pomeron/Reggeon are expected to generate additional tracks in the final state. In what follows we will explore these characteristics, as well as the presence of additional cutoffs, in order to separate the different mechanisms and reduce the background for the exclusive and double diffractive dimuon production. We will focus our analysis in the acceptances of the CMS and LHCb detectors, which probe complementary kinematical ranges. However, our study can be easily extended for the ATLAS acceptance.

\begin{center}
\begin{table}[t]
\begin{tabular}{ l|c|c|c|c|c|c|c|c }
\hline 
\hline
Cut\textbackslash{}Process & $\mathbb{P}\mathbb{P}$ & $\mathbb{PR}$ + $\mathbb{RP}$ & $\mathbb{RR}$ &
\textbf{DD} & $\mathbb{P}\boldsymbol{p}$ & $\mathbb{R}\boldsymbol{p}$ & \textbf{SD} & $\gamma\gamma$\tabularnewline
\hline 
\hline
\begin{tabular}{l}
{\footnotesize{}No cut}\tabularnewline
\end{tabular} & {\footnotesize{} 31.0 } & {\footnotesize{} 27.0 } &  {\footnotesize{} 6.1 }&
{\footnotesize{} \textbf{64.1} } & {\footnotesize{} 694.0 } & {\footnotesize{} 425.0 }& {\footnotesize{} \textbf{1119.0} } & {\footnotesize{} 7101.1 }\tabularnewline
\hline 
\begin{tabular}{l}
{\footnotesize{} 1. $p_{T}\left(\mu^{\pm}\right)>0.4\,\mbox{GeV}$}\tabularnewline
\end{tabular} & {\footnotesize{}28.6} & {\footnotesize{}23.9} & {\footnotesize{}4.5} &
{\footnotesize{}\textbf{57.3}} & {\footnotesize{}616.4} & {\footnotesize{}310.3} &{\footnotesize{}\textbf{926.7}} & {\footnotesize{}2601.3}\tabularnewline
\hline 
\begin{tabular}{l}
{\footnotesize{} 2. Inv. mass range $1.0 \le M_{\mu^+\mu^-} \le 20$ GeV}\tabularnewline
\end{tabular} & {\footnotesize{}23.3} & {\footnotesize{}19.3}  & {\footnotesize{}2.6} &
{\footnotesize{}\textbf{45.2}} & {\footnotesize{}499.6} & {\footnotesize{}189.5} & {\footnotesize{}\textbf{689.1}} & {\footnotesize{}1531.1}\tabularnewline
\hline 
\begin{tabular}{l}
{\footnotesize{} 3. $p_{T}^{2}\left(\mu^{+}\mu^{-}\right)<2\,\mbox{GeV}^{2}$}\tabularnewline
\end{tabular} & {\footnotesize{}16.5} & {\footnotesize{}13.0}  & {\footnotesize{}1.5} &
{\footnotesize{}\textbf{31.0}} & {\footnotesize{}236.1} & {\footnotesize{}82.2} &{\footnotesize{}\textbf{318.2}} & {\footnotesize{}1529.5}\tabularnewline
\hline 

\begin{tabular}{l}
{\scriptsize{} 4. $\eta$ in the CMS acceptance}\\
{\scriptsize{\,\,\,\,\,\,\,\,\,$\eta$ in the LHCb acceptance}}\tabularnewline
\end{tabular} & {\footnotesize{}}%
\begin{tabular}{l}
{\footnotesize{}5.7}\tabularnewline
{\footnotesize{}1.7}\tabularnewline
\end{tabular} & {\footnotesize{}}%
\begin{tabular}{l}
{\footnotesize{}3.4}\tabularnewline
{\footnotesize{}1.4}\tabularnewline
\end{tabular}  & {\footnotesize{}}%
\begin{tabular}{l}
{\footnotesize{}0.8}\tabularnewline
{\footnotesize{}0.1}\tabularnewline
\end{tabular} & {\footnotesize{}}%
\begin{tabular}{l}
{\footnotesize{}\textbf{9.8}}\tabularnewline
{\footnotesize{}\textbf{3.2}}\tabularnewline
\end{tabular} & {\footnotesize{}}%
\begin{tabular}{l}
{\footnotesize{}66.6}\tabularnewline
{\footnotesize{}20.8}\tabularnewline
\end{tabular} & {\footnotesize{}}%
\begin{tabular}{l}
{\footnotesize{}46.9}\tabularnewline
{\footnotesize{}6.2}\tabularnewline
\end{tabular} & {\footnotesize{}}%
\begin{tabular}{l}
{\footnotesize{}\textbf{113.5}}\tabularnewline
{\footnotesize{}\textbf{27.0}}\tabularnewline
\end{tabular} & {\footnotesize{}}%
\begin{tabular}{l}
{\footnotesize{}775.3}\tabularnewline
{\footnotesize{}46.6}\tabularnewline
\end{tabular}\tabularnewline
\hline 
{\footnotesize{}}%
\begin{tabular}{l}
{\scriptsize{} 5. Exclusivity: CMS}\tabularnewline 
{\scriptsize{}\,\,\,\,\,\,\, Exclusivity: Backward and forward LHCb}\tabularnewline
\end{tabular}  & {\footnotesize{}}%
\begin{tabular}{l}
{\footnotesize{}1.3} \tabularnewline	
{\footnotesize{}0.1}\tabularnewline
\end{tabular} & {\footnotesize{}}%
\begin{tabular}{l}
{\footnotesize{}1.2}\tabularnewline
{\footnotesize{}0.1}\tabularnewline
\end{tabular}  & {\footnotesize{}}%
\begin{tabular}{l}
{\footnotesize{}0.5}\tabularnewline
{\footnotesize{}0.01}\tabularnewline
\end{tabular} & {\footnotesize{}}%
\begin{tabular}{l}
{\footnotesize{}\textbf{3.0}}\tabularnewline
{\footnotesize{}\textbf{0.2}}\tabularnewline
\end{tabular} & {\footnotesize{}}%
\begin{tabular}{l}
{\footnotesize{}16.4}\tabularnewline
{\footnotesize{}0.9}\tabularnewline
\end{tabular} & {\footnotesize{}}%
\begin{tabular}{l}
{\footnotesize{}12.3}\tabularnewline
{\footnotesize{}0.6}\tabularnewline
\end{tabular} & {\footnotesize{}}%
\begin{tabular}{l}
{\footnotesize{}\textbf{28.7}}\tabularnewline
{\footnotesize{}\textbf{1.4}}\tabularnewline
\end{tabular} & {\footnotesize{}}%
\begin{tabular}{l}
{\footnotesize{}775.3}\tabularnewline
{\footnotesize{}46.6}\tabularnewline
\end{tabular}\tabularnewline
\hline
\hline
\end{tabular}
\caption{Predictions for the total cross sections in pb associated to the different mechanisms for the dimuon production for events with $p_T^2 <  2.0$ GeV$^2$ after the implementation of the different cuts discussed in the text. }
\label{table:cutslowpt}
\end{table}
\end{center}

\begin{center}
\begin{figure}[t]
\hspace*{-2.05em}\includegraphics[width=0.33\textwidth]{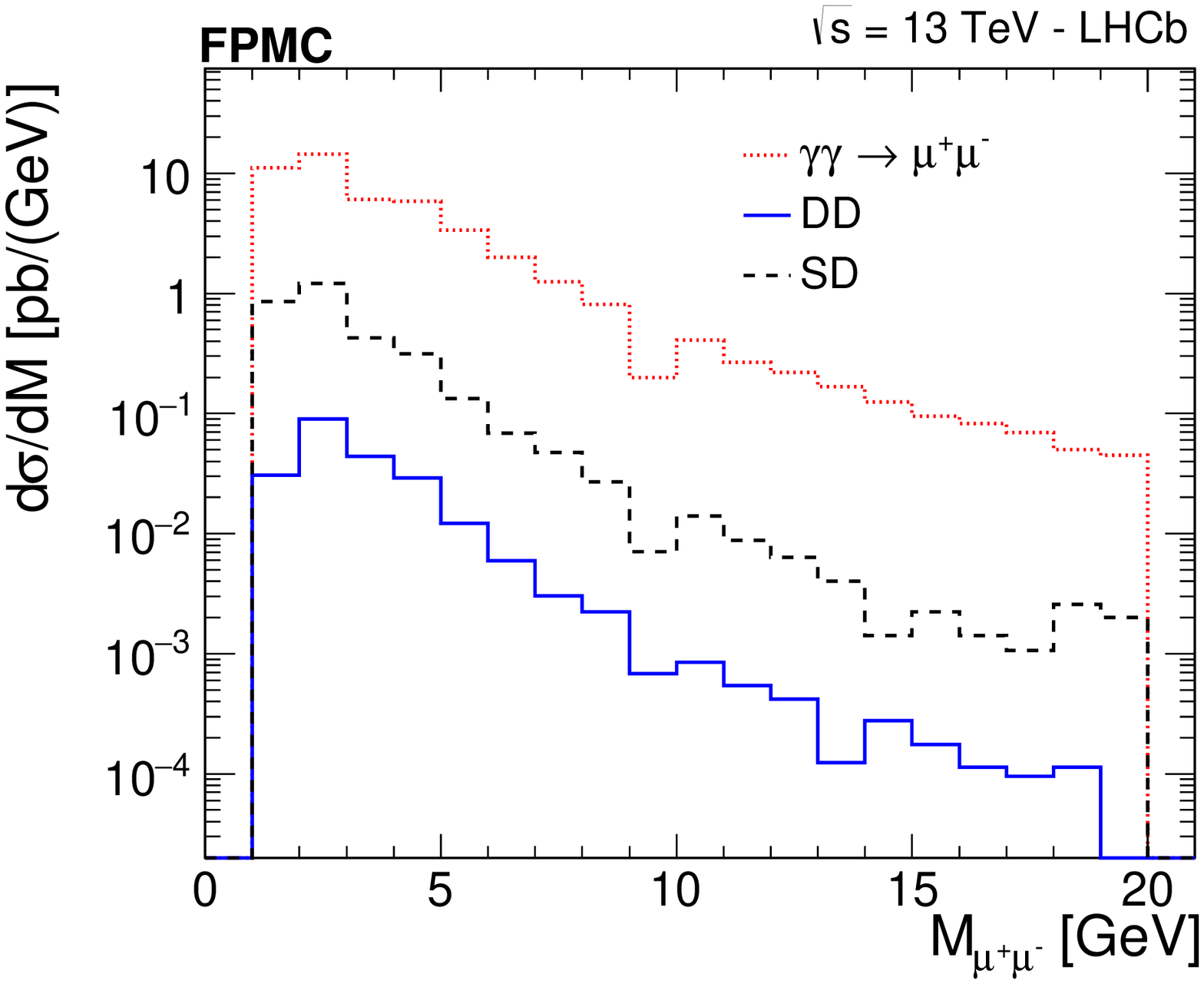}
\hspace*{-0.10em}\includegraphics[width=0.33\textwidth]{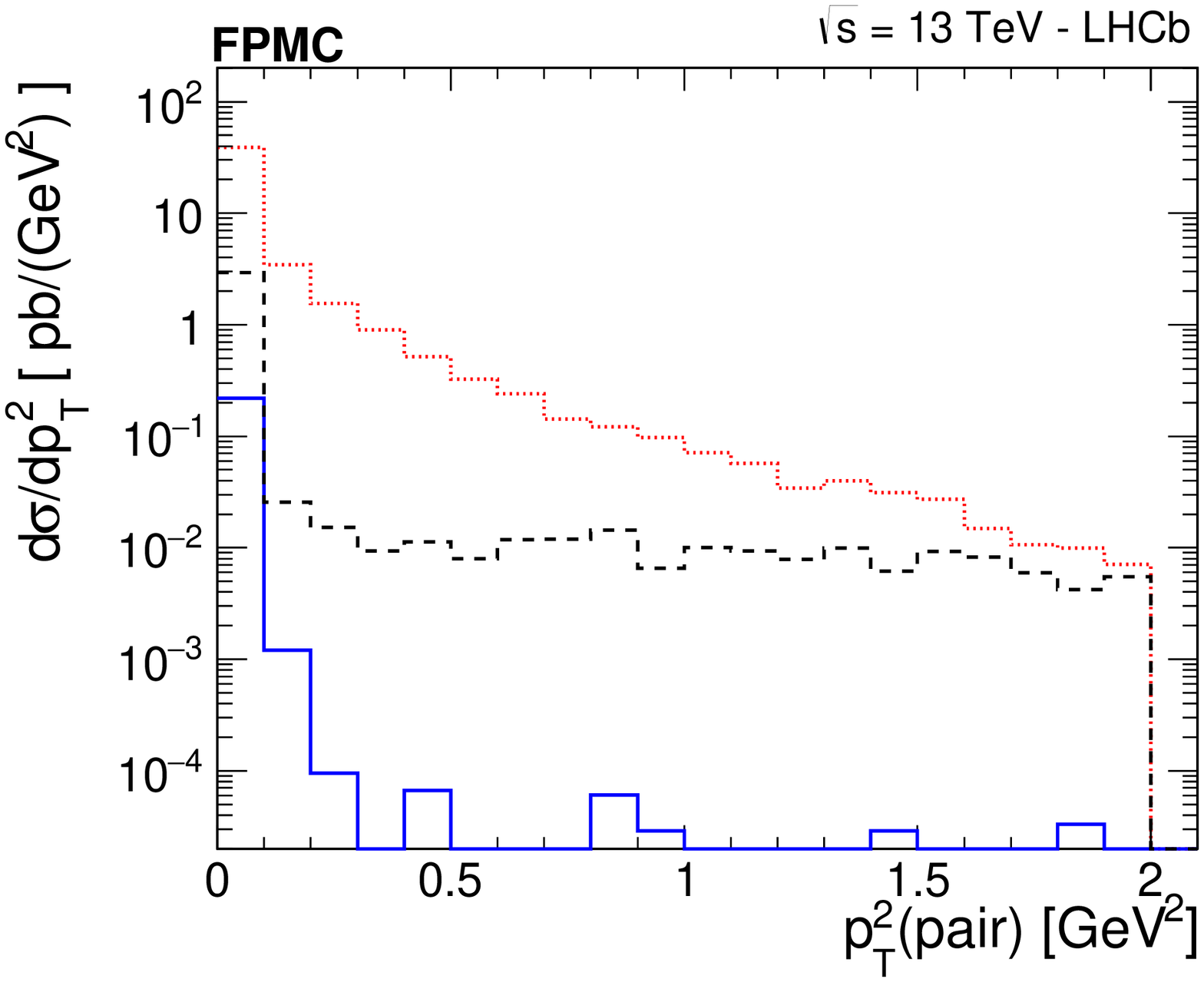}
\hspace*{-0.10em}\includegraphics[width=0.33\textwidth]{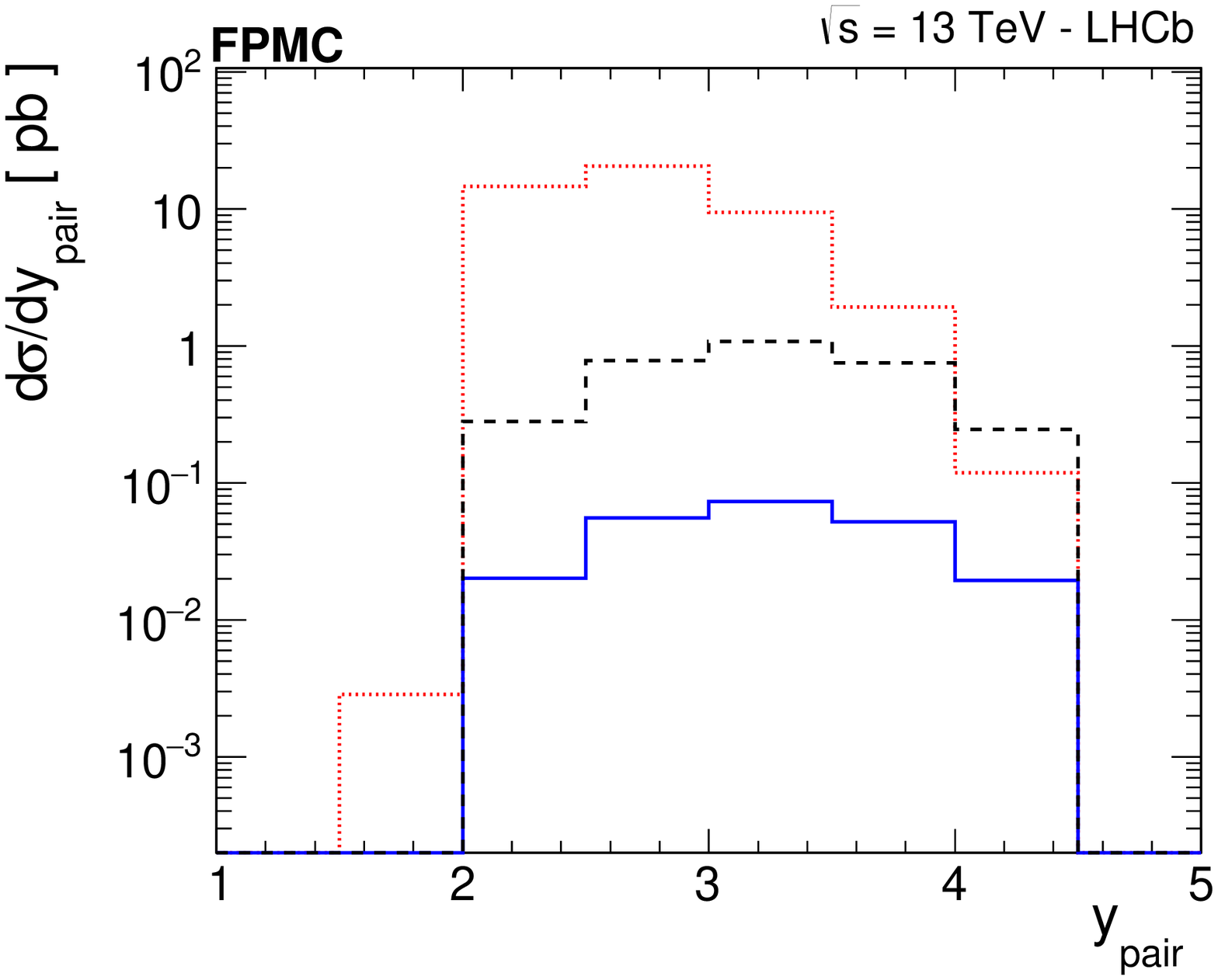}
\hspace*{-2.05em}\includegraphics[width=0.33\textwidth]{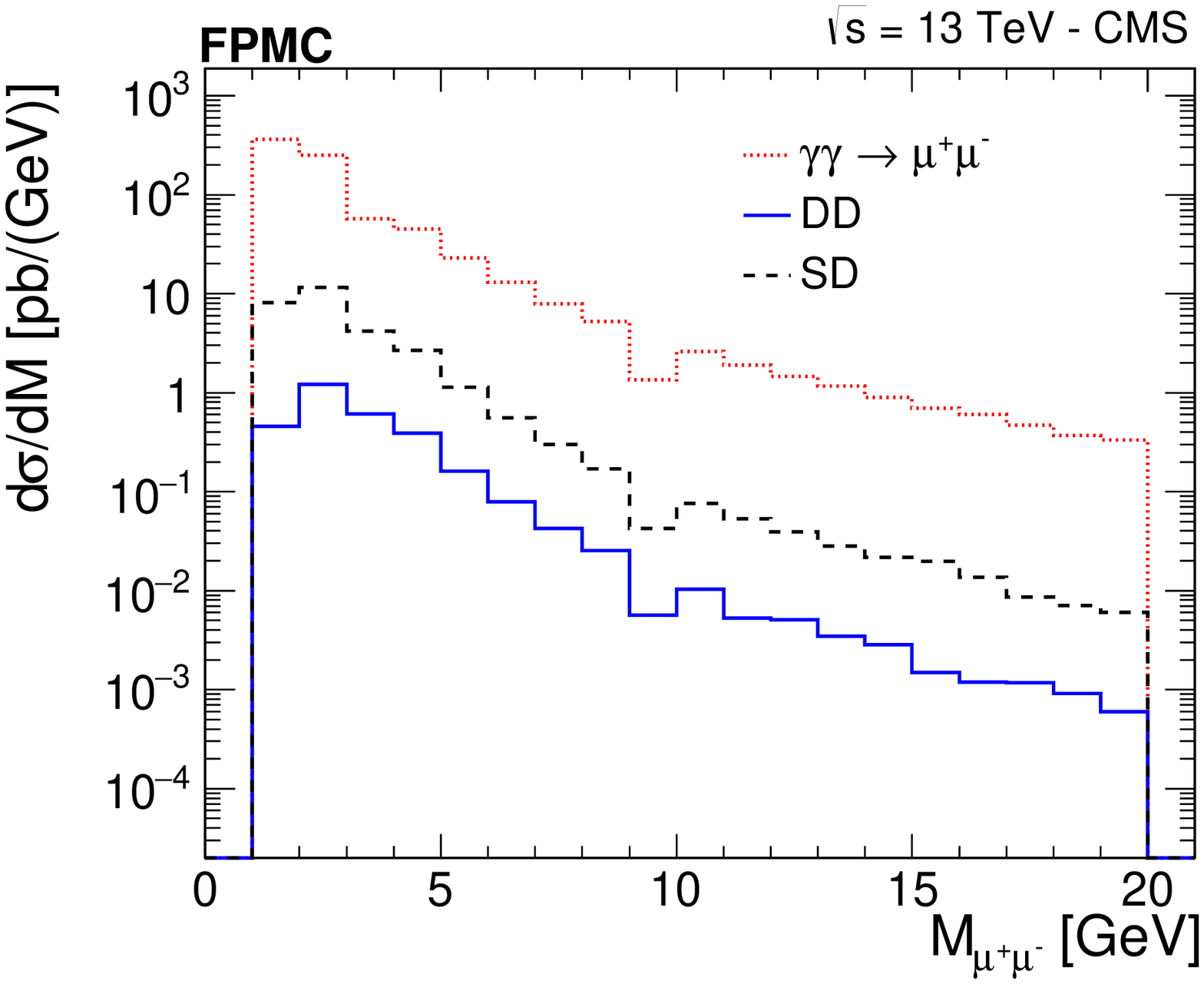}
\hspace*{-0.10em}\includegraphics[width=0.33\textwidth]{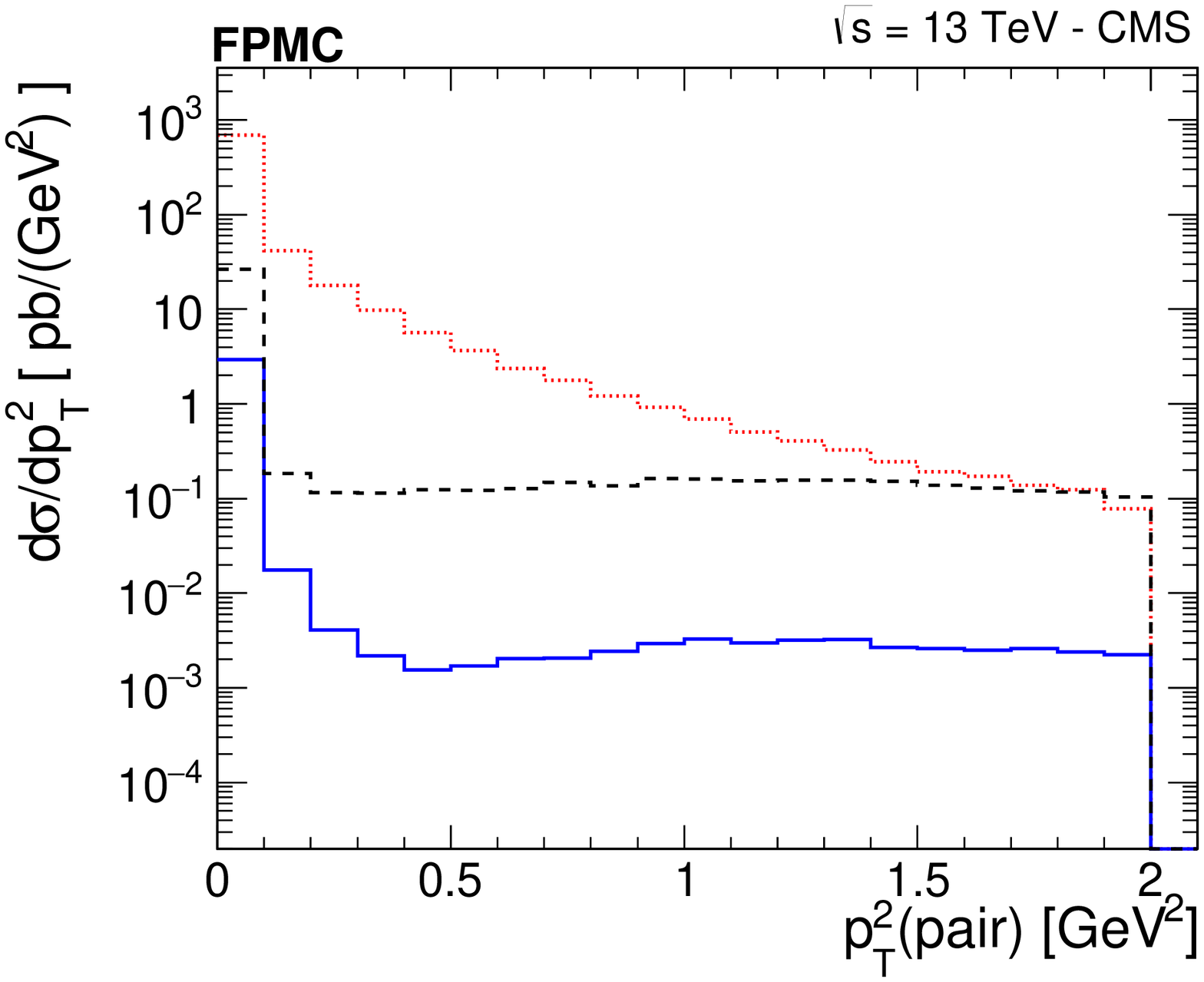}
\hspace*{-0.10em}\includegraphics[width=0.33\textwidth]{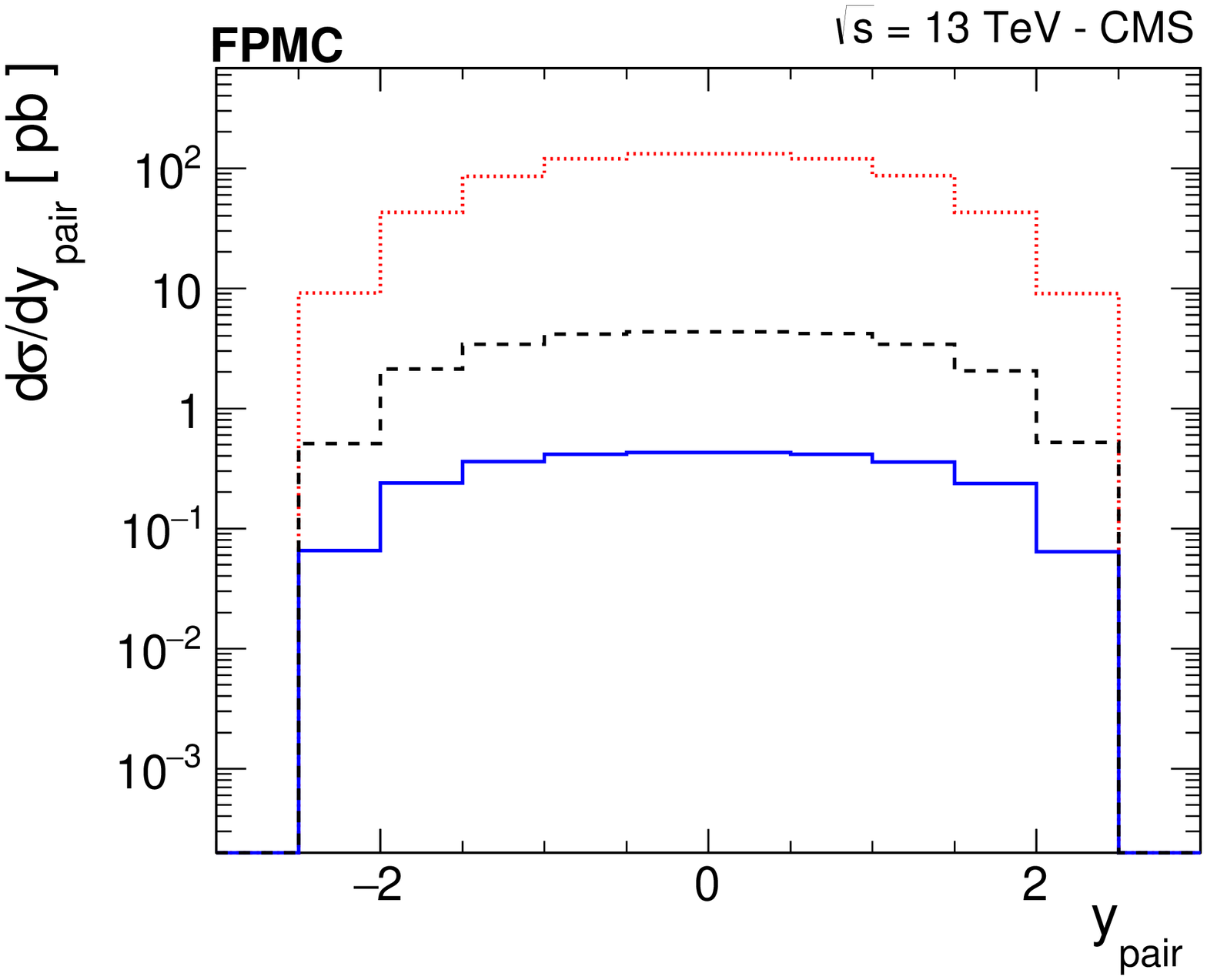}
\caption{Comparison between the predictions for the exclusive, single and double diffractive dimuon production after the implementation of the elastic selection criteria discussed in the text. }
\label{fig:cutslowpt}
\end{figure}
\end{center}

Initially let's analyze in more detail the small $p_T^2$ region, where we expect the dominance of the exclusive $\mu^+ \mu^-$ production by $\gamma \gamma$ interactions. In order to separate these events we will assume a set of requirements, usually denoted by {\it elastic selection} in the literature. In particular, we will assume that:
\begin{enumerate}
\item The individual transverse momenta of the muons is larger than a minimum: $p_{T}\left(\mu^{\pm}\right) > 0.4 \mbox{GeV}$;
\item The invariant mass of the dimuon system is in the range:  $1.0 \le M_{\mu^+ \mu^-} \le 20.0$ GeV. We have rejected the events with invariant mass in the range of the low mass and quarkonia resonances. In particular, the bands of rejection were: (a) For Low mass resonances: $M_{\mu^+ \mu^-} < 1.5\,\mbox{GeV}$, (b) For $J/\psi$:    $2.796\, \mbox{GeV} < M_{\mu^+ \mu^-} < 3.196\, \mbox{GeV}$, (c) For $\Psi(2S)$:  
$3.586\, \mbox{GeV}/< M_{\mu^+ \mu^-} < 3.786\, \mbox{GeV}$, and (d) For $\Upsilon$:  
$9.0\, \mbox{GeV} < M_{\mu^+ \mu^-} < 10.6\, \mbox{GeV}$;
\item The squared transverse momentum of the pair is smaller than a maximum: $ p_{T}^{2}\left(\mu^{+}\mu^{-}\right) < 2.0 \,\mbox{GeV}^{2}$. 
\item The pseudorapidities of the muons are in following ranges: (a) $\left|\eta\left(\mu^{\pm}\right)\right|<2.5$ in the case of the CMS detector, and (b)  {$2.0<\eta\left(\mu^{\pm}\right)<4.5$ in the case of the LHCb one;}
\item The event is exclusive, with only the pair of muons is present in the acceptance detector region. {For the CMS experimet we select events with  0 extra tracks with
$p_{T}>0.2\, \mbox{GeV}$ in the aceptance detector region. On the other hand,  in the case of the LHCb experiment, we select events with only the 
pair of muons in the range $2.0<\eta(\mu^{\pm})<4.5$  and 0 extra tracks with $p_{T}>0.1\, \mbox{GeV}$ in the regions $-3.5<\eta<-1.5, \,\, 
1.5<\eta<5.0$ and $p_{T}>0.5\, \mbox{GeV}$ in the regions $-8.0<\eta<-5.5, \,\, 5.5<\eta<8.0$ }.
\end{enumerate}
The impact of each one of these cuts on the total cross sections associated to the different mechanisms for the dimuon production is presented in Table \ref{table:cutslowpt}, where the values in each new line represent the predictions obtained after the inclusion of one additional cut. The final values, obtained after the implementation of the five selection criteria discussed above, are presented in the last two lines of the Table. We have that the diffractive contributions are strongly reduced, with the SD one being now a factor $\approx 27\,(31)$ smaller than the exclusive contribution in the CMS (LHCb) acceptance. The impact of the elastic selection on the invariant mass, squared transverse momentum and rapidity distributions are shown in
Fig. \ref{fig:cutslowpt}, where we present in the upper (lower) panels the predictions for the LHCb (CMS) detector. As expected from our results for the total cross sections, we have that the exclusive $\mu^+ \mu^-$ production by $\gamma \gamma$ interactions is dominant in the invariant mass range considered. In the particular case of the LHCb experiment (upper panels), we have that it also dominates the $p_T^2$ distribution and the SD contribution only becomes important at very forward rapidities. Our results indicate that the final $\mu^+ \mu^-$  events at LHCb can be used to probe of the exclusive dimuon production. In the case of the CMS experiment (lower panels), 
the exclusive production also is dominant, with the SD contribution only being important for $p_T^2 \approx 2$ GeV$^2$. It is important to emphasize that if the tagging of the {\it two} outgoing hadrons is implemented in the future,  the background associated to the SD events vanish. As a consequence, the final events will be a clean probe of the exclusive dimuon production.

\begin{center}
\begin{table}[t]
\begin{tabular}{ l|c|c|c|c|c|c|c|c }
\hline 
\hline
Cut\textbackslash{}Process & $\mathbb{P}\mathbb{P}$ & $\mathbb{PR}$ + $\mathbb{RP}$ & $\mathbb{RR}$ &
\textbf{DD} & $\mathbb{P}\boldsymbol{p}$ & $\mathbb{R}\boldsymbol{p}$ & \textbf{SD} & $\gamma\gamma$\tabularnewline
\hline 
\hline
\begin{tabular}{l}
{\footnotesize{}No cut}\tabularnewline
\end{tabular} & {\footnotesize{} 31.0 } & {\footnotesize{} 27.0 } &  {\footnotesize{} 6.1 }&
{\footnotesize{} \textbf{64.1} } & {\footnotesize{} 694.0 } & {\footnotesize{} 425.0 }& {\footnotesize{} \textbf{1119.0} } & {\footnotesize{} 7101.1 }\tabularnewline
\hline 
\begin{tabular}{l}
{\footnotesize{} 1. $p_{T}\left(\mu^{\pm}\right)>0.4\,\mbox{GeV}$}\tabularnewline
\end{tabular} & {\footnotesize{}28.6} & {\footnotesize{}23.9} & {\footnotesize{}4.5} &
{\footnotesize{}\textbf{57.0}} & {\footnotesize{}616.4} & {\footnotesize{}310.3} &{\footnotesize{}\textbf{926.7}} & {\footnotesize{}2601.3}\tabularnewline
\hline 
\begin{tabular}{l}
{\footnotesize{} 2. Inv. mass range $1.0 \le M_{\mu^+\mu^-} \le 20$ GeV}\tabularnewline
\end{tabular} & {\footnotesize{}23.3} & {\footnotesize{}19.3}  & {\footnotesize{}2.6} &
{\footnotesize{}\textbf{45.2}} & {\footnotesize{}499.6} & {\footnotesize{}189.5} & {\footnotesize{}\textbf{689.1}} & {\footnotesize{}1531.1}\tabularnewline
\hline 
\begin{tabular}{l}
{\footnotesize{} 3. $p_{T}^{2}\left(\mu^{+}\mu^{-}\right)>2\,\mbox{GeV}^{2}$}\tabularnewline
\end{tabular} & {\footnotesize{}4.7} & {\footnotesize{}4.2}  & {\footnotesize{}0.6} &
{\footnotesize{} {\bf 9.6}} & {\footnotesize{}166.8} & {\footnotesize{}63.3} &{\footnotesize{}{\bf 230.1}} & {\footnotesize{}0.1}\tabularnewline
\hline 

\begin{tabular}{l}
{\scriptsize{} 4. $\eta$ in the CMS acceptance}\\
{\scriptsize{\,\,\,\,\,\,\,\,\,$\eta$ in the LHCb acceptance}}\tabularnewline
\end{tabular} & {\footnotesize{}}%
\begin{tabular}{l}
{\footnotesize{}2.2}\tabularnewline
{\footnotesize{}0.6}\tabularnewline
\end{tabular} & {\footnotesize{}}%
\begin{tabular}{l}
{\footnotesize{}1.7}\tabularnewline
{\footnotesize{}0.6}\tabularnewline
\end{tabular}  & {\footnotesize{}}%
\begin{tabular}{l}
{\footnotesize{}0.3}\tabularnewline
{\footnotesize{}0.1}\tabularnewline
\end{tabular} & {\footnotesize{}}%
\begin{tabular}{l}
{\footnotesize{}{\bf 4.3}}\tabularnewline
{\footnotesize{}{\bf 1.2}}\tabularnewline
\end{tabular} & {\footnotesize{}}%
\begin{tabular}{l}
{\footnotesize{}70.4}\tabularnewline
{\footnotesize{}17.6}\tabularnewline
\end{tabular} & {\footnotesize{}}%
\begin{tabular}{l}
{\footnotesize{}38.5}\tabularnewline
{\footnotesize{}5.8}\tabularnewline
\end{tabular} & {\footnotesize{}}%
\begin{tabular}{l}
{\footnotesize{}{\bf 108.9}}\tabularnewline
{\footnotesize{}{\bf 23.4}}\tabularnewline
\end{tabular} & {\footnotesize{}}%
\begin{tabular}{l}
{\footnotesize{}0.04}\tabularnewline
{\footnotesize{}0.005}\tabularnewline
\end{tabular}\tabularnewline
\hline 
{\footnotesize{}}%
\begin{tabular}{l}
{\scriptsize{} 5. Exclusivity: CMS}\tabularnewline 
{\scriptsize{}\,\,\,\,\,\,\, Exclusivity: Backward and forward LHCb}\tabularnewline
\end{tabular}  & {\footnotesize{}}%
\begin{tabular}{l}
{\footnotesize{}0.04} \tabularnewline	
{\footnotesize{}$8\times 10^{-4}$}\tabularnewline
\end{tabular} & {\footnotesize{}}%
\begin{tabular}{l}
{\footnotesize{}0.2}\tabularnewline
{\footnotesize{}0.002}\tabularnewline
\end{tabular}  & {\footnotesize{}}%
\begin{tabular}{l}
{\footnotesize{}0.08}\tabularnewline
{\footnotesize{}$5\times 10^{-4}$}\tabularnewline
\end{tabular} & {\footnotesize{}}%
\begin{tabular}{l}
{\footnotesize{}{\bf 0.3}}\tabularnewline
{\footnotesize{}{\bf 0.004}}\tabularnewline
\end{tabular} & {\footnotesize{}}%
\begin{tabular}{l}
{\footnotesize{}1.5}\tabularnewline
{\footnotesize{}0.01}\tabularnewline
\end{tabular} & {\footnotesize{}}%
\begin{tabular}{l}
{\footnotesize{}2.1}\tabularnewline
{\footnotesize{}0.01}\tabularnewline
\end{tabular} & {\footnotesize{}}%
\begin{tabular}{l}
{\footnotesize{}{\bf 3.6}}\tabularnewline
{\footnotesize{}{\bf 0.02}}\tabularnewline
\end{tabular} & {\footnotesize{}}%
\begin{tabular}{l}
{\footnotesize{}0.04}\tabularnewline
{\footnotesize{}0.005}\tabularnewline
\end{tabular}\tabularnewline
\hline
\hline
\end{tabular}
\caption{Predictions for the total cross sections in pb associated to the different mechanisms for the dimuon production for events with $p_T^2 >  2.0$ GeV$^2$ after the implementation of the different cuts discussed in the text. }
\label{table:cutshighpt}
\end{table}
\end{center}

\begin{center}
\begin{figure}[t]
\hspace*{-2.05em}\includegraphics[width=0.33\textwidth]{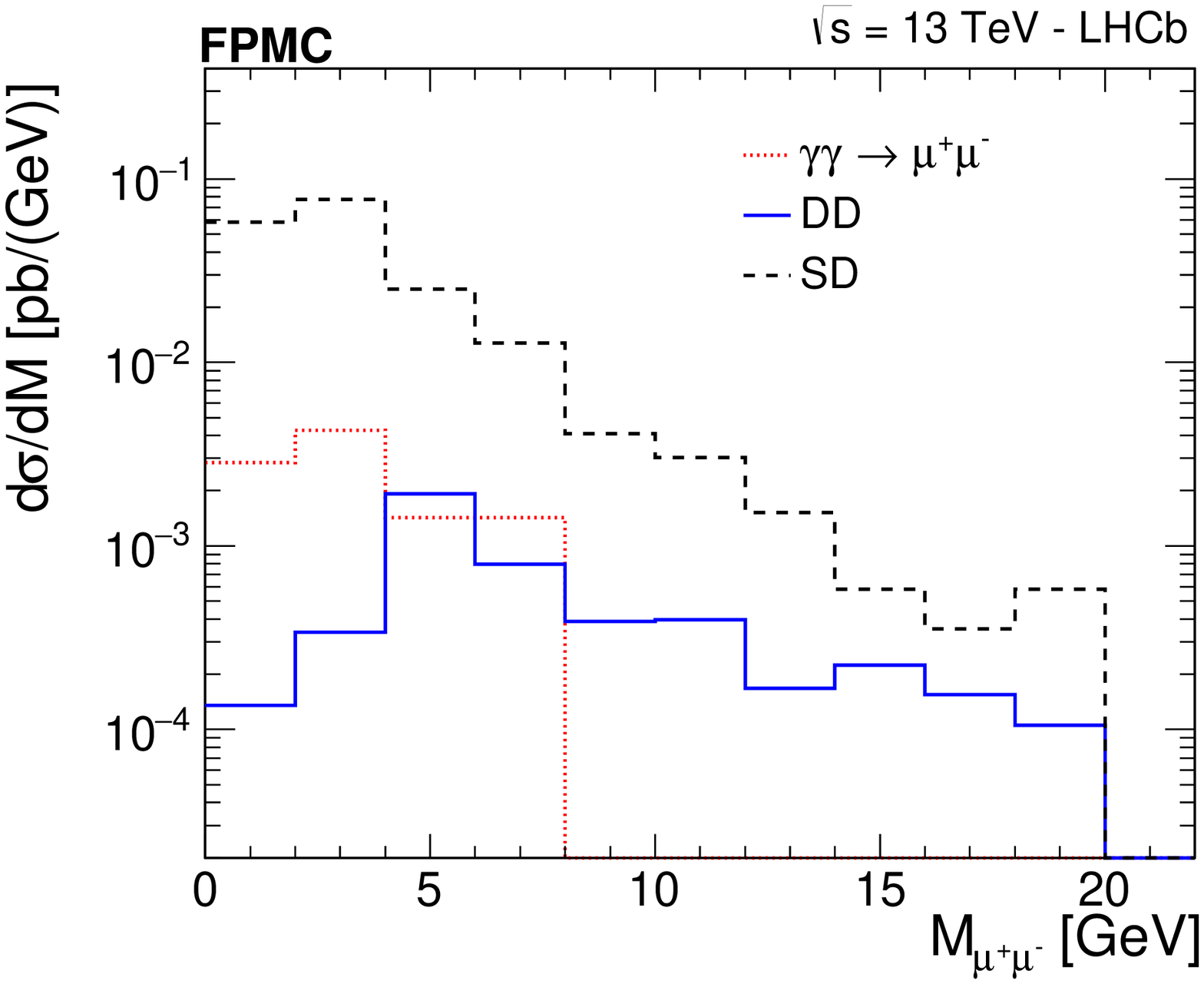}
\hspace*{-0.10em}\includegraphics[width=0.33\textwidth]{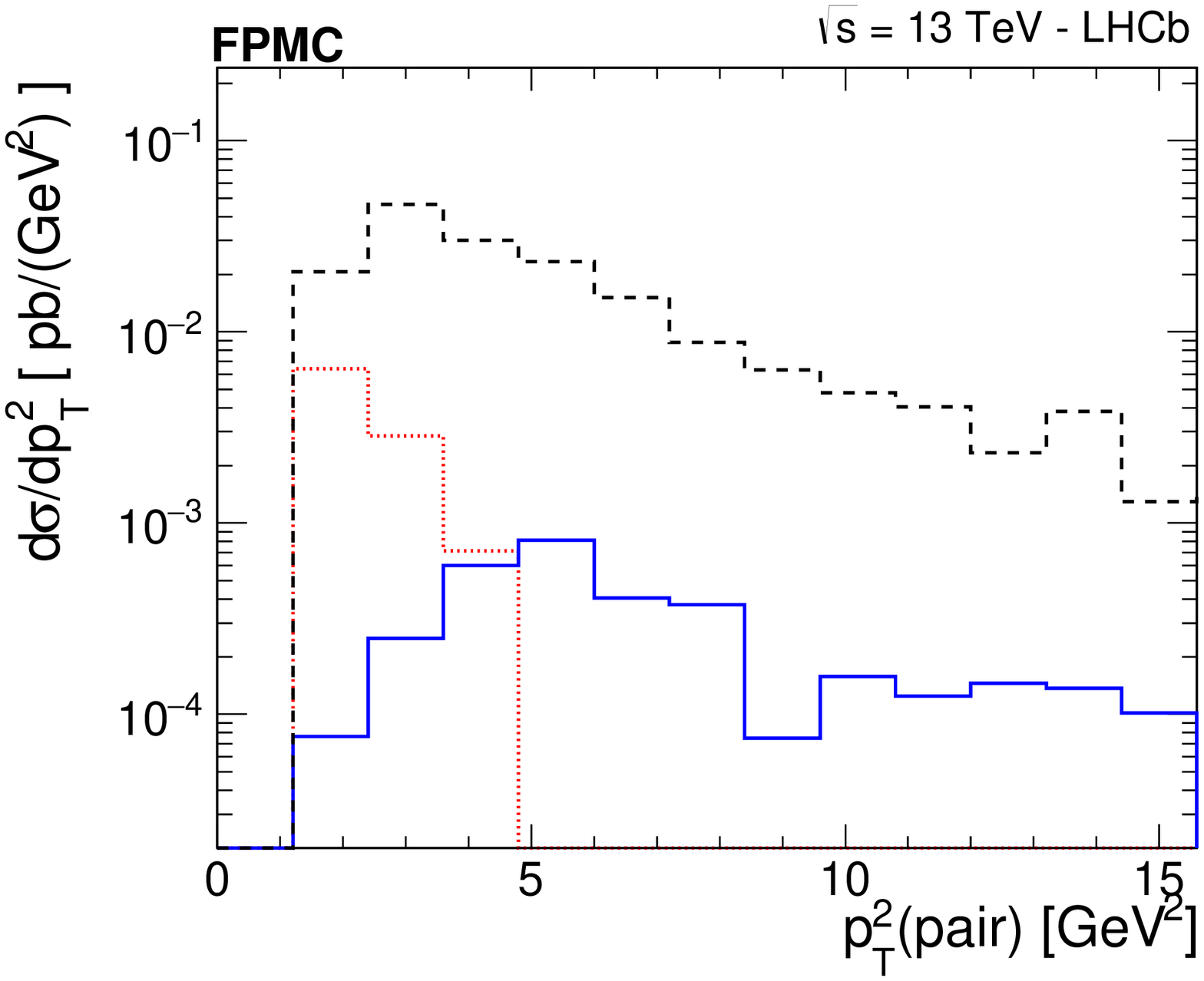}
\hspace*{-0.10em}\includegraphics[width=0.33\textwidth]{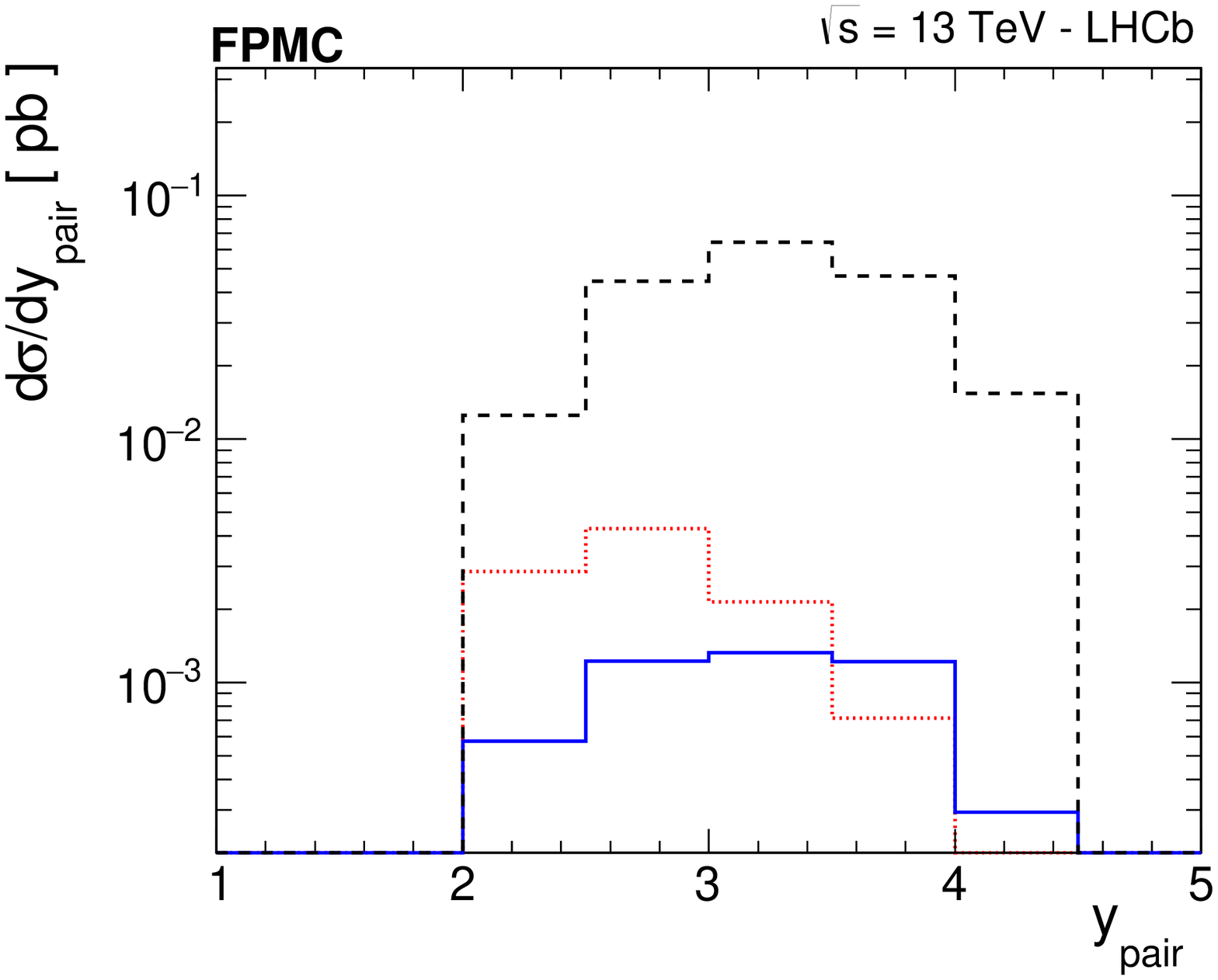}
\hspace*{-2.05em}\includegraphics[width=0.33\textwidth]{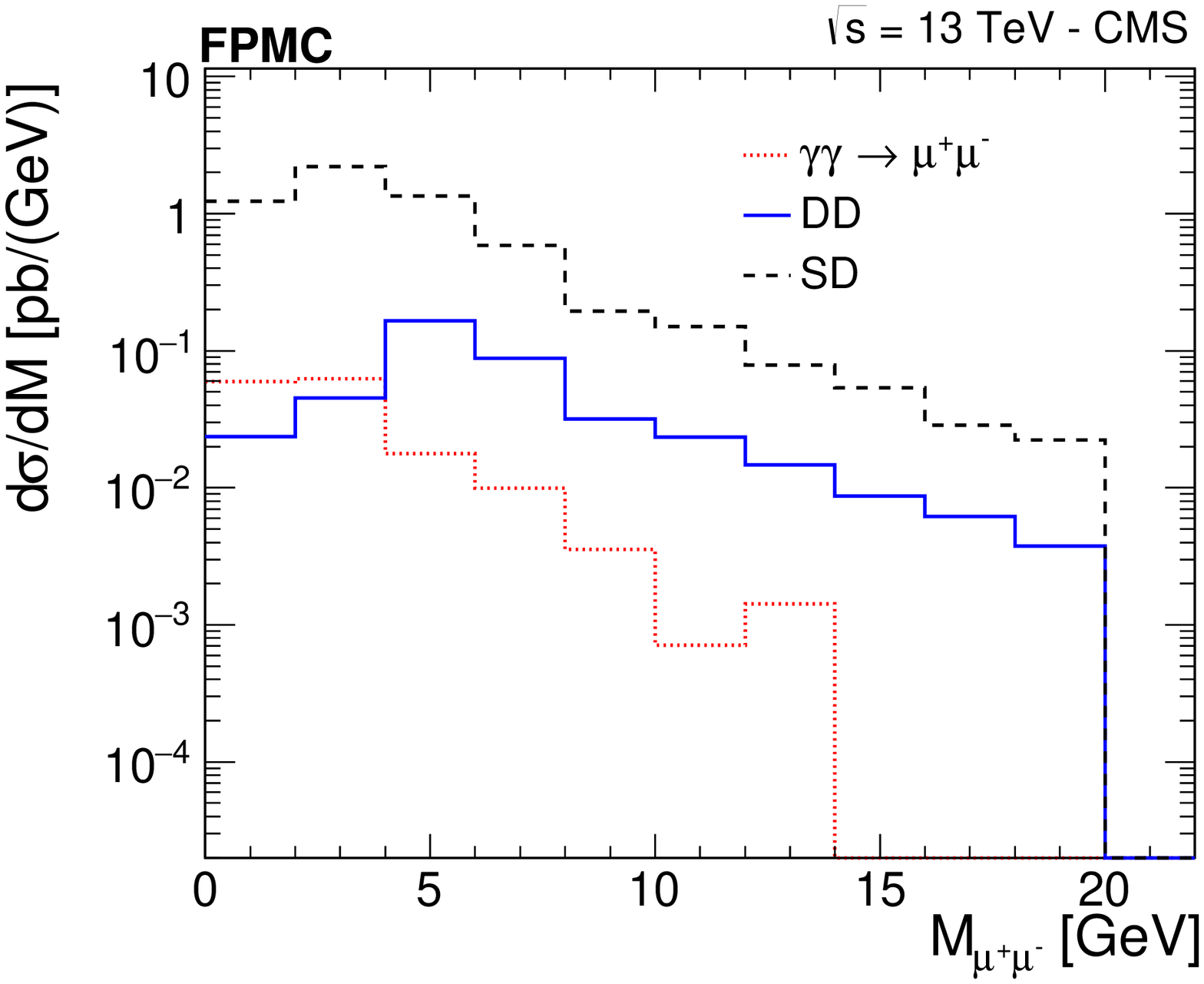}
\hspace*{-0.10em}\includegraphics[width=0.33\textwidth]{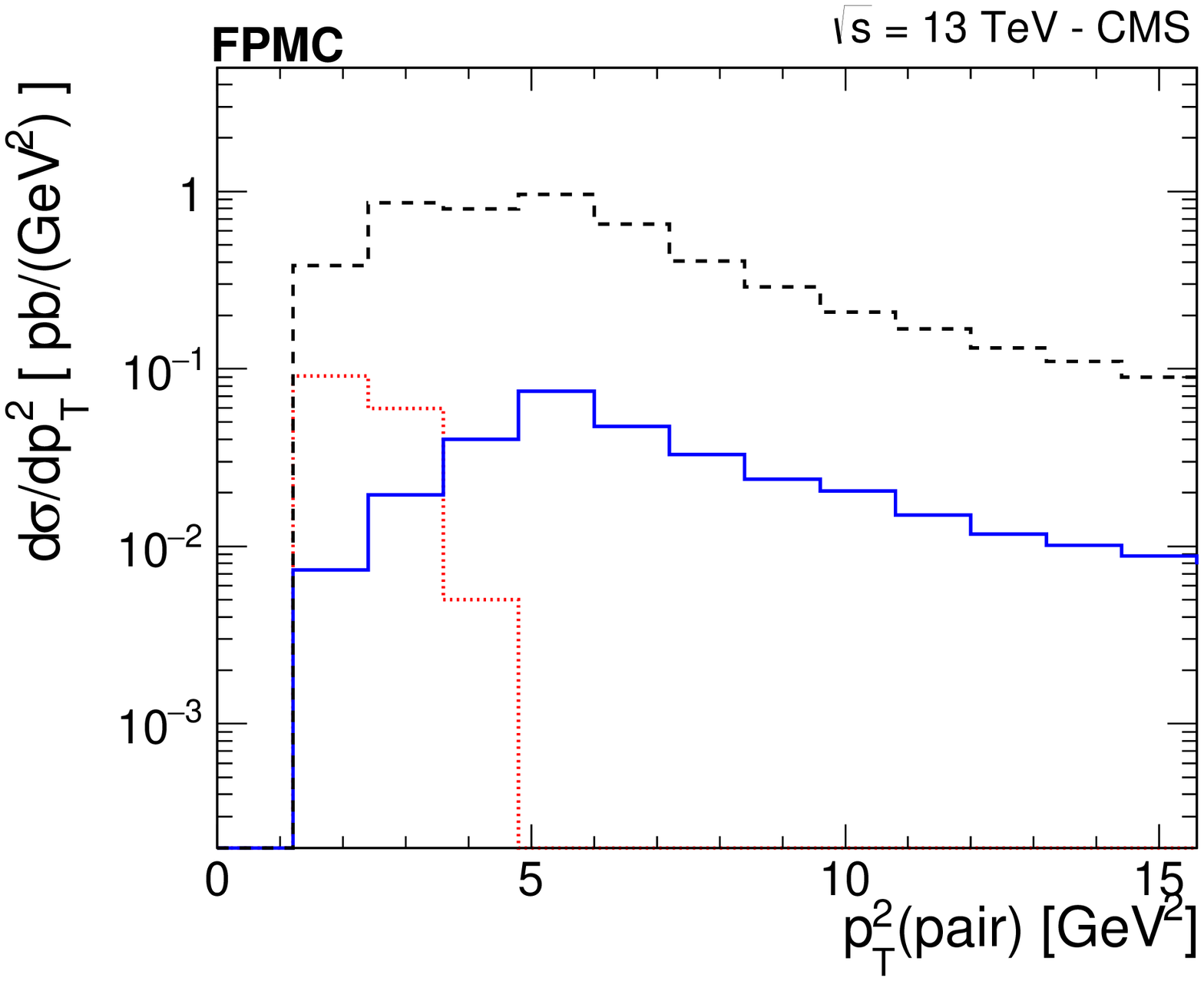}
\hspace*{-0.10em}\includegraphics[width=0.33\textwidth]{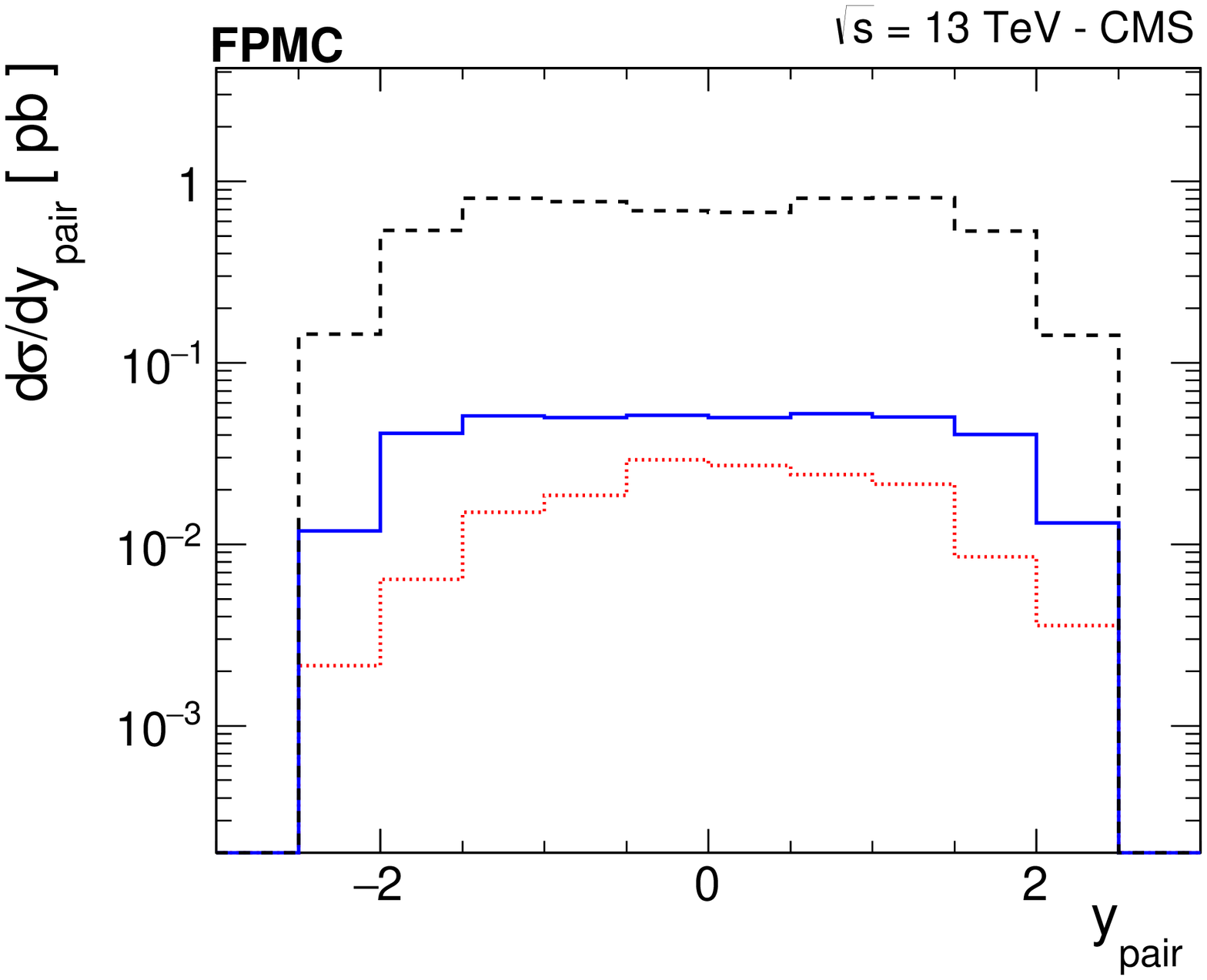}

\caption{Comparison between the predictions for the exclusive, single and double diffractive dimuon production considering events with $p_T^2 > 2$ GeV$^2$ and  implementation of the cuts discussed in the text.}
\label{fig:cuthighpt}
\end{figure}
\end{center}

Finally, let's analyze the possibility of constrain the diffractive processes using the events with large transverse momentum of the pair. We have implemented the same cuts discussed above, only  assuming now that the events should be characterized by $p_T^2 > 2$ GeV$^2$. The predictions for the total cross sections are presented in Table 
\ref{table:cutshighpt}. Our results indicate that the double diffractive events become dominant in the CMS acceptance and similar to the exclusive one at LHCb after the implementation of the cuts. The SD events are dominant for both detectors. The impact of the cuts on the distributions are presented in Fig. \ref{fig:cuthighpt}. In the case of the LHCb cuts (upper panels), we have that the diffractive mechanisms are dominant at large invariant mass and/or $p_T^2$. On the other hand, the rapidity distributions of the diffractive and exclusive mechanisms are similar. If the CMS cuts are assumed (lower panels), the diffractive contributions dominate all distributions. Our results indicate that the study of the high $p_T^2$ events can be useful to constrain the description of the diffractive dimuon production, as well its underlying assumptions as for example the modelling of the rapidity gap survival probability.

\section{Summary}
\label{conc}

The description of the exclusive and diffractive processes in hadronic collisions  is  a theme of intense debate in literature. Significant theoretical improvements have been achieved in recent years and abundant experimental data have been accumulated at the LHC. In particular, the study of the exclusive dimuon production have received a lot of attention, mainly motivated by the possibility of use this process as a luminosity monitor and as a probe of the photon distribution of the proton.  In this paper we have investigated in detail the dimuon production by $\gamma \gamma$ and diffractive interactions in $pp$ collisions at the LHC energy. Our goal was to determine the regions of dominance of these different mechanisms taking into account some realistic experimental requirements. Using the  Forward Physics Monte Carlo,  we have  performed a comprehensive analysis of the invariant mass, transverse momentum and rapidity distributions for the different processes. We have demonstrated that the separation of the small $p_T^2$ events allows to probe the exclusive $\mu^+ \mu^-$ production by $\gamma \gamma$ interactions with a small background associated to diffractive processes. On the other hand, the analysis of the large  $p_T^2$ events is an important probe of the description of the diffractive processes and underlying assumptions. Considering the large statistics of dimuons events, we strongly motivate a future experimental analysis of this process at large transverse momentum in order to advance in our understanding about Diffractive Physics.


\section*{Acknowledgements}
This work was  partially financed by the Brazilian funding
agencies CNPq, CAPES, FAPERJ, FAPERGS and  INCT-FNA (process number 
464898/2014-5).



\end{document}